\documentclass[conference,a4paper]{IEEEtran}

\usepackage{cite}

\usepackage[pdftex]{graphicx}
\graphicspath{{./graphics/}}

\usepackage[cmex10]{amsmath}
\usepackage{amssymb,amsfonts,relsize,bbm}		
\usepackage[utf8]{inputenc}
\usepackage[T1]{fontenc}
\usepackage[normalem]{ulem}  

\usepackage[utf8]{inputenc}
\usepackage{tikz}
\usetikzlibrary{shapes,arrows,chains,arrows,decorations.pathmorphing,backgrounds,positioning,fit,petri,calc,patterns}
\usetikzlibrary{matrix,positioning}
\usepackage{pgfplots}
\usepackage{caption,subcaption}
\usepgfplotslibrary{groupplots}
\pgfplotsset{compat=newest}

\usepackage[pscoord]{eso-pic}
\newcommand{\placetextbox}[3]{
\setbox0=\hbox{#3}
\AddToShipoutPictureFG*{
	\put(\LenToUnit{#1\paperwidth},\LenToUnit{#2\paperheight}){\vtop{{\null}\makebox[0pt][c]{#3}}}%
}
}

\usepackage{color}
\usepackage{algorithm,algorithmicx,algpseudocode}
\algdef{SE}[DOWHILE]{Do}{doWhile}{\algorithmicdo}[1]{\algorithmicwhile\ #1}		
\algrenewcommand\algorithmicforall{\textbf{foreach}}
\algrenewcommand\algorithmicrequire{\textbf{Input:}}
\algrenewcommand\algorithmicensure{\textbf{Output:}}

\newcommand{\gammaT}{\gamma^{\text{T}}}
\newcommand{\gammabarT}{\bar{\gamma}^{\text{T}}}

\DeclareMathOperator{\indicator}{\mathbbm{1}}

\newcommand{\second}{$2^{\mathrm{nd}}$}  
\newcommand{\third}{$3^{\mathrm{rd}}$}  
\newcommand{\Pmax}{P^{\text{max}}}  
\newcommand{\acir}{\lambda}
\newcommand{\acirf}{\lambda_{|f'-f|}}

\newcommand{\Ab}{\mathbf{A}}
\newcommand{\Xb}{\mathbf{X}}
\newcommand{\xb}{\mathbf{x}}
\newcommand{\Yb}{\mathbf{Y}}
\newcommand{\Zb}{\mathbf{Z}}
\newcommand{\Pb}{\mathbf{P}}
\newcommand{\Hb}{\mathbf{H}}

\newcommand{\Wb}{\mathbf{W}}
\newcommand{\Ms}{\mathcal{M}}
\newcommand{\Ns}{\mathcal{N}}
\newcommand{\Fs}{\mathcal{F}}
\newcommand{\Ts}{\mathcal{T}}
\newcommand{\Rs}{\mathcal{R}}

\newcommand{\Ss}{\mathcal{S}}

\newcommand{\Ps}{\mathcal{P}}

\newcommand{\Js}{\mathcal{J}}

\newcommand{\Us}{\mathcal{U}}

\newcommand{\nTx}{N^{\mathrm{Tx}}}
\newcommand{\nRx}{N^{\mathrm{Rx}}}

\newcommand{\davg}{$d_{\text{avg}}$}

\newcommand{\figref}[1]{Fig.~\ref{#1}}
\newcommand{\comments}[1]{}
\newcommand{\arrivalTime}{t^\mathrm{gen}_m}

\DeclareMathOperator*{\argmin}{arg\,min}
\IEEEoverridecommandlockouts

\usepackage{glossaries}
\newacronym{ACI}{ACI}{adjacent channel interference}
\newacronym{ACIR}{ACI}{adjacent channel interference ratio}
\newacronym{AOI}{AoI}{age of information}
\newacronym{BLP}{BLP}{Boolean linear programming}
\newacronym{BS}{BS}{base station}
\newacronym{CAMs}{CAMs}{cooperative awareness messages}
\newacronym{CCI}{CCI}{co-channel interference}
\newacronym{CSI}{CSI}{channel state information}
\newacronym{CSMA}{CSMA}{carrier-sense multiple access}
\newacronym{D2D}{D2D}{device-to-device}
\newacronym{DENMs}{DENMs}{decentralized environmental notification messages}
\newacronym{ETSI}{ETSI}{European Telecommunications Standards Institute}
\newacronym{HARQ}{HARQ}{hybrid automatic repeat request}
\newacronym{ITS-S}{ITS-S}{intelligent transport system stations}
\newacronym{LP}{LP}{linear programming}
\newacronym{LTE}{LTE}{long term evolution}
\newacronym{MAC}{MAC}{medium access control}
\newacronym{MBLP}{MBLP}{mixed Boolean linear programming}
\newacronym{PA}{PA}{power amplifier}
\newacronym{QoS}{QoS}{quality of service}
\newacronym{RB}{RB}{resource block}
\newacronym{RSU}{RSU}{road side unit}
\newacronym{SINR}{SINR}{signal to interference and noise power ratio}
\newacronym{STR}{STR}{Signal to total received power ratio}
\newacronym{VUE}{VUE}{vehicular user equipment}
\newacronym{V2V}{V2V}{vehicle-to-vehicle}
\newacronym{V2X}{V2X}{vehicle-to-everything}
\newacronym{RRM}{RRM}{radio resource management}

\begin{document}
\title{Radio Resource Management for V2V Multihop Communication Considering Adjacent Channel Interference}
\author{Anver~Hisham,
	Erik~G.~Str\"{o}m,~\IEEEmembership{Senior~Member,~IEEE},
	and~Fredrik~Br\"annstr\"om,~\IEEEmembership{Member,~IEEE} \\
	\thanks{Anver Hisham, Erik G. Str\"{o}m, and Fredrik Br\"annstr\"om are with the Communication Systems Group, Department of Electrical Engineering, Chalmers University of Technology, SE-412 96 Gothenburg, Sweden. E-mail: \{anver, erik.strom, fredrik.brannstrom\}@chalmers.se}
	\thanks{The research was funded and performed in the framework of the H2020 project 5GCAR co-funded by the EU. The authors would like to acknowledge the contributions of their colleagues. The views expressed are those of the authors and do not necessarily represent the project. }\\
}

\maketitle	
\placetextbox{0.83}{0.98}{\Large \parbox{4cm}{ Draft 8\\ \today}}%

\begin{abstract}
	This paper investigate joint scheduling and power control for V2V multicast allowing multihop communication. The effects of both co-channel interference and adjacent channel interference are considered. First we solve the problem with the objective of maximizing the throughput and connectivity of vehicles in the network. Then extend the same problem formulation to include the objective of minimizing the latency and the average \gls{AOI}, which is the age of the latest received message. In order to account for the fairness, we also show the problem formulation to maximize the worst-case throughput and connectivity. All the problems are formulated as mixed Boolean linear programming problems, which allows computation of optimal solutions. Furthermore, we consider the error probability of a link failure in all the problem formulations and accommodate the probability requirements for satisfying a certain throughput/connectivity/latency/\gls{AOI}. In order to support a large V2V network, a clustering algorithm is proposed whose computational complexity scale well with the network size. To handle the case of zero channel information at the scheduler, a multihop distributed scheduling scheme is proposed.
\end{abstract}

\section{Introduction}

\subsection{Motivation}
\Gls{V2V} communications have drawn great attention due to its ability to improve traffic safety and efficiency. \Gls{V2V} communication can reduce accidents by broadcasting up-to-date local and emergency information. To this end, both periodic and event-driven messages are conveyed. 

Periodic messages are broadcasted by all vehicles to inform neighbors about their current state, i.e., position, speed, heading, acceleration, etc\comments{for driver assistance and awareness}, while event-driven messages are sent when an emergency situation has occurred. To this end, \gls{ETSI} is standardizing both \gls{CAMs} for periodic messages, and \gls{DENMs} for aperiodic messages. \gls{CAMs} are sent with frequency 2--100\,Hz with proposed latency requirements of 3--100\,ms, depending upon the application \cite{3gpp22.186}. \gls{DENMs} are used to alert vehicles of a detected event, and the transmission can be repeated and persists as long as the event is present \cite{3gpp302.637-3}. However, both periodic and aperiodic messages in V2V are broadcast and localized in its nature, i.e., since they are to facilitate cooperation between vehicles in close proximity. Due to the safety-critical nature of the communication, latency and reliability requirements are stringent for V2V safety related communication. The most demanding applications require a combination of low latency and ultra-reliable communication.


Latency is typically defined as end-to-end delay a message experiences and reliability is the probability that the latency is not exceeding the latency requirement. In the \gls{V2V} setting, the latency metric is mainly applicable to DENM messages, as consecutive DENM messages are, in general, not interrelated. CAM messages, on the other hand, carry periodic state updates and consecutive messages are clearly interrelated. For such traffic, the \gls{AOI} metric is more applicable. The \gls{AOI}, at a particular receiver, is the age of the latest received message. The \gls{AOI} is therefore a non-negative random process, and performance can be expressed in terms of the statistics of, e.g., the peak or the time-average of the process~\cite{devassy.durisi.ferrante.simeone.uysal:2019}. \Gls{AOI} has been studied in the \gls{V2V} context in, e.g., \cite{Sanjit1}.  

To control the latency or \gls{AOI}, it is important to have control over the packet error probability.
The packet error probability depends on the \gls{SINR}, which in turn depends upon the received interference power. There are two main types of interference: \gls{CCI} and \gls{ACI}. \Gls{CCI} is the cross-talk between transmitters scheduled in the same time-frequency slot, while \gls{ACI} is the interference due to the leakage of transmit power outside the intended frequency slot. Therefore, ACI affect transmissions scheduled in the same timeslot, but in different frequency slots. ACI is mainly caused by the nonlinearities of the \gls{PA} in the transmitter.  Advanced methods have been developed to linearize the \gls{PA} \cite{Mohammadi1,Lavrador1,Kenington1,Jessica1}, however, the clipping effect of the \gls{PA} cannot be avoided, which results in ACI. 

Typically, ACI is negligible compared to CCI when the ACI interferers are not very much closer to the receiver compared to the desired transmitter. This is usually the case in cellular downlink/uplink communication. However, in V2V communication, distances to transmitters or interferers can be highly varying. Furthermore, the penetration loss by blocking vehicles (which increases with respect to the carrier frequency \cite{Zaidi1}) is significant for V2V communication at 5.9\,GHz\cite{Abbas2,Taimoor,Meireles1,He1}. This implies that the received power ratio from a nearby to a far-away transmitter is high, especially when there are many blocking vehicles. Consequently, the desired signal could be weak compared to strong interfering signals (resulting in a so-called near-far situation), and ACI can be a significant problem \cite{Anver_Access1,AnverArchive_Journal2}. Therefore, the effects of ACI should also be considered along with CCI while designing \gls{RRM} schemes (i.e., scheduling and power control) for  direct V2V communication.

The penetration loss due to blocking vehicles or buildings might prevent connectivity even over short distances. To enable connectivity in these cases requires multihop (relaying) communication---either through the fixed infrastructure, e.g., via an uplink/core network/downlink, or via a number of V2V direct links. There are pros and cons with each arrangement. However, the latter case is the only option when vehicles are outside coverage of the fixed infrastructure. Even when inside coverage, if the source and destination are relative close to each other, it might be more resource efficient to multihop via vehicles than via the fixed infrastructure~\cite{fodor.dahlman.mildh.parkvall.reider.miklos.turanyi:2012}. Moreover, lower range of V2V communication allows spectrum re-usage within a small area, while the large range of fixed infrastructure limits its possibilities. 


\subsection{State of the Art}
As already mentioned, ACI is not a significant problem in cellular communication, therefore most of the existing literature focuses solely on mitigating CCI alone without considering ACI \cite{v2vsch1,v2vsch2,v2vsch3}. Still, the impact of ACI for cellular uplink communication and \gls{D2D} communication has been analyzed in \cite{Li1} and \cite{Albasry1} respectively. The impact of ACI on 802.11b/g/n/ac has also been broadly studied \cite{aciw1,aciw2,aciw3}. All these studies generally conclude that ACI causes outage and performance degradation. Additionally, for \gls{V2V} communication with \gls{CSMA} \gls{MAC} layer, ACI can cause a potential transmitter to falsely assume that the channel is busy resulting in deferring transmissions \cite{Campolo1,Campolo2}.

Multihop communication in V2V has also gained much attention recently, e.g., see \cite{Yilin1,Kassen1,Saket1}, and references therein. In  \cite{Yilin1}, an optimization problem is formulated to maximize the throughput and minimize latency using multihop routing. A theoretical analysis on the packet error probability bounds for multihop communication has been done in \cite{Giovanni1}, and the authors conclude that 1-hop communication is beneficial when vehicle density is low. In \cite{Subramanian1}, authors approach multihop scheduling from a graph theoretic point of view and propose novel algorithms. Minimizing average \gls{AOI} in vehicular networks has also captured attention and widely studied recently \cite{Sanjit1,Andrea1,Zhou1,Abd1}.

However, none of the above studies on scheduling or power control considers the effects of ACI. Furthermore, there is no study which combines multihop scheduling and \gls{AOI}. Our previous studies \cite{Anver_Access1, AnverArchive_Journal2} try to find efficient scheduling and power control algorithms while taking into account the effects of ACI. In this paper, we generalize our previous work in mainly four directions: 1) allowing for multihop communication, 2) considering \gls{AOI} as a performance metric, 3) introducing clustering to ensure scalability, 4) proposing a distributed scheduling algorithm.



\subsection{Contributions}
We make the following contributions in this paper:
\begin{enumerate}
	\item The joint scheduling and power control problem to maximize the average/worst-case throughput and connectivity of a V2V network are formulated as \gls{MBLP} problems. 
	\item Similar problem formulations are done to maximize connectivity with certain requirements on latency and \gls{AOI}.
	\item Due to the high computational complexity in finding optimal scheduling and power values for large networks, we propose a clustering based algorithm which reduces computational complexity to ensure scalability. 
	\item A low-complexity, cluster-based distributed scheduling algorithm is proposed, in which a \gls{VUE} is required to know only its position index, network size, and cluster size.
	\item In all the problem formulations and proposed algorithms in this paper, we allow multihop communication and optimize considering the effects of both \gls{CCI} and \gls{ACI}.
\end{enumerate}

\subsection{Notation and Outline} 
We use the following notation throughout the paper. Lowercase and uppercase letters, e.g., $x$ and $X$, represent scalars. Lowercase boldface letters, e.g., $\xb$, represent a vector where $x_{i}$ is the $i^{\mathrm{th}}$ element of $x$. Uppercase boldface letters, e.g., $\mathbf{X}$, denote matrices where $X_{i,j}$ indicates the $(i,j)^\mathrm{th}$ element. The notation $|\Xb|$ denote the number of elements in matrix $\Xb$. Calligraphic letters, e.g., $\mathcal{X}$, represent sets, $|\mathcal{X}|$ denote its cardinality, and $\emptyset$ denotes the empty set. We use $a\bmod b$ for the remainder of $a$ when divided by $b$. The notation $\lceil \cdot \rceil$, and $\lfloor \cdot \rfloor$, $\lfloor \cdot \rceil$ represents ceil, floor, and round operations, respectively. The Boolean OR, AND and NOT operations are denoted by $\vee$, $\wedge$, and $\neg$, respectively. The indicator function $\indicator\{\textit{statement}\}$ is equal to 1 if \textit{statement} is true and 0 otherwise.

The paper is organized as follows. Section \ref{sec:SystemModel} presents the system model and Section \ref{sec:joint} shows the problem formulations for various objectives. To address the scalability for large networks, we present a clustering based \gls{RRM} algorithm in Section \ref{sec:Clustering}. A distributed algorithm for resource allocation is presented in Section \ref{sec:CDS} and computational complexity of all algorithms are computed in Section \ref{sec:ComputationalComplexity}. Finally, the simulation results are presented in Sections \ref{sec:PerformanceEvaluation} and conclusion in Section \ref{sec:Conclusions}.


\section{System Model}  \label{sec:SystemModel}

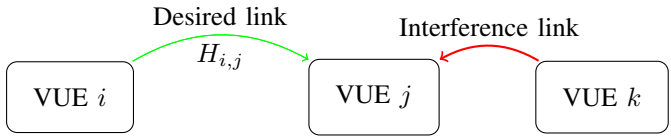
\begin{figure}[t]
	\centering	
\begin{tikzpicture}[rec/.style={rounded corners,inner sep=10pt,draw}]
\node(VUEi) at (0,0) [rec]{VUE $i$} ;
\node(VUEj) at (4,0) [rec]{VUE $j$} ;
\node(VUEk) at (7,0) [rec]{VUE $k$} ;
\draw[bend left,->,color=green] (VUEi) to node[above,black]{Desired link} (VUEj);
\draw[bend right,->,color=red,thick] (VUEk) to node[above,black]{Interference link} (VUEj);
\draw[bend left,->,color=green] (VUEi) to node[below,black]{$H_{i,j}$} (VUEj);
\end{tikzpicture}
	\caption{System model}\label{drawing:SystemModel}
\end{figure}

We consider a network of $N$ VUEs in the set 
\begin{equation}
 \Ns \triangleq \{0,1,\cdots,N-1\}   
\end{equation}
We consider multicast communication, which includes unicast and broadcast as special cases. We define $\Rs_i\subset \Ns$ as the set of \emph{intended receivers} to VUE~$i$. That is, VUE $i$ wish to transmit its messages to the VUEs in $\Rs_i$. Clearly,
$|\Rs_i|= 1$ and $\Rs_i = \Ns\setminus \{i\}$ implies that VUE~$i$ use unicast communication and broadcast communication, respectively. 

In total, $M$ messages are generated in the network during the scheduling interval. The available time-frequency resources are divided into $T$ timeslots and $F$ frequency slots. A time-frequency slot is called a \gls{RB} and is denoted as $(f,t)$, where 
\begin{align}
    f\in\Fs\triangleq\{0, 1, \ldots, F-1\},\\
    t\in\Ss\triangleq\{0, 1, \ldots, T-1\}.
\end{align}
For simplicity, we assume that the message
\begin{equation}
    m\in\Ms\triangleq\{0, 1, \ldots, M-1\}
\end{equation}
can be transmitted in a single RB. If a message is too large to fit into an RB, then it has to be scheduled in multiple RBs as explained in Appendix \ref{Appendix:SupportingLargePayloads}.
The maximum transmit power of a VUE is~$\Pmax$. 

In general, we denote the transmitting, receiving, and interfering VUEs with $i$, $j$, and $k$, respectively, as illustrated in \figref{drawing:SystemModel}. The link $(i,j)$ indicates the link from VUE $i$ to VUE $j$. 

The parameter $H_{i,j}$ is the average channel power gain from VUE $i$ to VUE $j$. Hence, $H_{i,j}$ takes into account pathloss, penetration loss and large-scale fading between VUE $i$ and VUE $j$. We assume that $H_{i,j}$ is fixed during the scheduling interval (i.e., for $T$ timeslot durations) and known to the scheduling and power control algorithms. For practical V2V channels, this implies that the scheduling interval is on the order of 100~ms. The small-scale fading (fast fading) distribution is assumed to be known. However, the fast fading \emph{realization} is not assumed to be known, as this would require a potentially very large \gls{CSI} measurement and feedback overhead.


Suppose VUE~$i$ is transmitting in RB $(f,t)$ and VUE~$k$ in RB $(f',t')$. If $t'\neq t$, then there will be no interference, since timeslots are assumed to be orthogonal. If $t' = t$, then there will \gls{CCI} if $f'=f$ and \gls{ACI} if $f'\neq f$. In this paper, we consider the effects of both \gls{CCI} and \gls{ACI}. It can be shown that the transmitted message will be received with an error probability less than $\epsilon$ if the \gls{SINR} (as defined in~\eqref{definition:SINR} below) is equal or larger than a certain threshold $\gammaT$. The threshold can computed for any given $\epsilon$ and small-scale fading distribution~\cite[Lemma 1]{Wanlu2016}.



In general, scheduling and power control is done by a controller. As mentioned above, we assume that large-scale channel parameters (i.e., pathloss, shadowing, and penetration loss) are slowly varying compared to the scheduling interval $T$ and that the controller has access to this slowly varying \gls{CSI} for all relevant VUE pairs. 
A base station (BS), \gls{ITS-S}, or a specially assigned VUE can act as the controller. In this paper, we will consider the case when the network has a single controller, multiple controllers, and when each VUE acts as its own controller.


\begin{table}[t] 
	\caption{Key Mathematical Symbols}\label{table_notation}
	\renewcommand{\arraystretch}{1.3}
	\begin{tabular}[t]{cl}
		\hline		Symbol  &   Definition  \\ 	
		\hline    
		\underline{Parameters}  &  \\
		$N$ & Number of VUEs \\
		$F$ & Number of frequency slots \\
		$T$ & Number of timeslots \\
		$M$ & Total number of messages to transmit \\
		$H_{i,j}$ & Average channel power gain from VUE $i$  to VUE $j$ \\
		$\acir_r$ & ACI from any frequency slot $f$ to frequency slot $f \pm r$ \\ 
		$\sigma^2$ & Noise power in an RB \\		
		$\gammaT$ & \parbox[t]{6.5cm}{SINR threshold to declare a link is successful}\\
		$\Pmax$ & Maximum transmit power of a VUE \\
		$\Ms_{i}$ & Set of messages generated by VUE $i$ \\
		$\arrivalTime$ & Generation time of message $m$ \\[0.1cm]
		$\Omega_{i,m,t}$ & \parbox{7cm}{Indicate if VUE $i$ can transmit the message $m$ at the earliest timeslot $t$}   \vspace*{0.2cm}  \\
		\underline{Variables}  &  \\
		$\Rs_i$ & Set of receivers for Tx-VUE $i$ \\
		$P_{i,f,t}$ & Transmit power of VUE $i$ in an RB in timeslot $t$ \\
		$S_{i,j,f,t}$ & Received power by VUE $j$ from VUE $i$ in RB $(f,t)$ \\
		$R_{j,f,t}$ & Total received power by VUE $j$ in RB $(f,t)$ \\
		$X_{i,m,f,t}$ &  \parbox[t]{6.5cm}{Indicate if VUE $i$ is scheduled to transmit message $m$ in RB $(f,t)$} \\
        $Y_{i,j,f,t}$ & Indicate if link $(i,j)$ is successful in RB $(f,t)$ \\
		$W_{j,m,t}$ & \parbox[t]{6.5cm}{Indicate if VUE $j$ receives message $m$ during timeslot $t$} \\
		$Z_{i,j}$ & Indicate if link $(i,j)$ is connected or not \\
		$A_{i,j,t}$ & \gls{AOI} of the link $(i,j)$ during timeslot $t$  \\  
		$\tau_{j,m}$  & Latency of message $m$ upon reception by VUE $j$  \\
		$C$ & Number of clusters \\
		$G$ & Number of groups \\
		$\nTx$ & Number Tx-VUEs in a group \\
		$\Ts^{(c,g)}$ & Set of Tx-VUEs in group $(c,g)$ \\
		$\Rs^{(c,g)}$ & Set of Rx-VUEs in group $(c,g)$ \\
		$\Ss_g$ & Set of timeslots for group $g$ \\
		\hline
	\end{tabular}
\end{table}

\section{Joint Scheduling and Power Control} \label{sec:joint}

In this section, joint scheduling and power control problem to maximize various objectives are formulated as an \gls{MBLP} problem. Note that, all Boolean operations (like AND, OR, ... etc) can be translated into linear operations with Boolean variables as explained in Appendix \ref{Appendix:MathematicalBackground}. Key mathematical symbols are listed in Table \ref{table_notation}. 

\subsection{Variables and Constraints Formulations}

In this section, we will define a number of variables and constraints that are indexed by $i$, $j$, $m$, $f$, and $t$. If not explicitly stated otherwise, the definitions are valid for $i\in\Ns$, $j\in\Ns$, $m\in\Ms$, $f\in\Fs$, and $t\in\Ss$.
 
\subsubsection{Message generation time}
We assume that a VUE generates a message $m$ at time $\arrivalTime \in \mathbb{R}$, and that it is available for transmission on or after time $\arrivalTime + t^\mathrm{d}$, where $t^\mathrm{d} \in \mathbb{R}^+$ is the minimum time delay between message generation and transmission. Both $\arrivalTime$ and $t^\mathrm{d}$ are assumed to be measured in terms of number of timeslot durations.

We define the parameter $\Omega_{i,m,t} \in \{0,1\}$ to indicate if VUE~$i$ generates the message $m$ and it is available for transmission at the earliest timeslot $t$, i.e.,
\begin{equation}
\Omega_{i,m,t} \triangleq 
\begin{cases}
1, & \parbox{5cm}{\raggedright if VUE $i$ generates message $m$ during timeslot $\lfloor t-t^\mathrm{d} \rfloor$} \\
0, & \parbox{5cm}{\raggedright otherwise.} \\
\end{cases}
\end{equation}
That is, if $\Omega_{i,m,t}=1$, then the message $m$ is available for transmission on or after the timeslot $t$. We assume that the message arrivals are deterministic, hence, $\Omega_{i,m,t}$ is known in the optimization problems. We denote the set of messages generated by VUE~$i$ as
\begin{equation}
    \Ms_{i} \triangleq \{m\in\Ms: \Omega_{i,m,t}=1, t\in\Ss\}.  
\end{equation}


\subsubsection{Scheduling constraints}
The elements of the scheduling matrix $\Xb\in\{0,1\}^{N\times N\times F\times T}$ are the variables
\begin{equation}
X_{i,m,f,t} \triangleq  
\begin{cases}
1, & \parbox{5cm}{\raggedright if VUE $i$ is scheduled to transmit message $m$ in RB $(f,t)$ } \\
0, & \parbox{5cm}{\raggedright otherwise.} \\
\end{cases}  
\end{equation}

A VUE can transmit at most one message in an RB. Hence, since
\begin{equation}
    \tilde{X}_{i,f,t} = \sum_{m\in\Ms} X_{i,m,f,t}
\end{equation}
indicates if VUE $i$ is transmitting in RB $(f,t)$, the constraint is
\begin{equation}
    \tilde{X}_{i,f,t}  \leq 1. \label{constr:Xift}
\end{equation}

\subsubsection{Transmit power constraints}
The transmit power matrix is denoted $\Pb\in [0,\Pmax]^{N\times F\times T}$, where $P_{i,f,t}$ is the transmit power of VUE $i$ in RB $(f,t)$. The variable $P_{i,f,t}$ is constrained by the maximum transmit power $\Pmax$ of a VUE:
\begin{equation}
    \sum_{f\in\Fs} P_{i,f,t} \leq \Pmax    \label{constr:power_total}   
\end{equation}
Furthermore, $P_{i,f,t}$ is also constrained by scheduling as
\begin{equation}  
0 \leq P_{i,f,t} \leq \Pmax \tilde{X}_{i,f,t}.     \label{constr:PConstrainedByX}	
\end{equation}

\subsubsection{\gls{SINR} constraints}

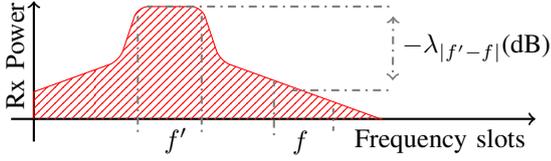
\begin{figure}[t]
       \centering
\tikzstyle{rArea} = [pattern=north east lines, draw=red,pattern color=red]
\tikzstyle{lStyle} = [dash dot,thick,black!50]

\begin{tikzpicture}
\begin{scope}[yshift=-4cm,scale = 1.2]{Receiver}

\draw[thick,->] (-.25,0) -- (5.5,0) node[anchor=north east]{Frequency slots};
\draw[thick,->] (0,-.25) -- (0,1.3);
\node[anchor=west,rotate=90] at (-.2cm,-0) {Rx Power};

\draw[rArea,xshift=0cm] (0,0) -- (0,0.3)  {[rounded corners] -- ++(20:1) -- ++(0.2,0.6) node(fp1){} --  ++(0.7,0) node(fp2){} --  ++(0.2,-0.6)}  --node(f1){} ++(-20:1.9);

\draw[lStyle,dash phase=40pt] (fp1) --  ++(0,-1.4) node[xshift=0.5cm,yshift=-0.1cm,black](){$f'$};
\draw[lStyle,dash phase=40pt] (fp2) --  ++(0,-1.4);

\draw[lStyle,dash phase=40pt] ($(f1)+(-0.3,0.1)$) node(f1new){} --  ++(0,-0.6) node[xshift=0.35cm,yshift=-0.1cm,black](){$f$};
\draw[lStyle,dash phase=40pt] ($(f1new)+(0.65,-0.3)$) --  ++(0,-0.25);

\draw[lStyle,dash phase=40pt] (fp1) --  ++(2.8,0) node(lambdaTop){};
\draw[lStyle,dash phase=40pt] ($(f1new)+(0.2,-0.1)$) --  ++(1.1,0) node(lambdaBottom){};
\draw[lStyle,dash phase=40pt,<->] (lambdaTop) -- node[xshift=1.1cm, black](){$-\lambda_{|f'-f|}$(dB)} (lambdaBottom);

\end{scope}
\end{tikzpicture}
       \caption{Received power (dBm) when an interferer is transmitting in frequency slot $f'$}\label{drawing:ACIR}
\end{figure}

Suppose VUE~$i$ transmits a message to VUE~$j$ in RB $(f,t)$. The desired signal power at VUE~$j$ is
\begin{equation}
    S_{i,j,f,t} = P_{i,f,t} H_{i,j},   \label{definition:S}
\end{equation}
and the total received signal power (desired plus interference) is
\begin{equation}
    R_{j,f,t} = \sum_{f'\in\Fs}  \sum_{k\in\Ns}  P_{k,f',t}H_{k,j}\acirf, \label{definition:I}   
\end{equation}
where $\acir_r$ is the \gls{ACIR} from a frequency slot $f$ to frequency slot $f \pm r$ \cite[section 17.9]{DahlmanParkvallSkold}. Therefore, $\acirf$ is the inverse-\gls{ACIR} from frequency slot $f'$ to $f$, see \figref{drawing:ACIR}. In other words, $\acirf$ is the ratio of the received interference power in frequency slot $f$ to the received interference power in frequency slot $f'$ when the interfering VUE is transmitting in frequency slot $f'$. Note that when $f'= f$, then the interference is CCI instead of ACI. Therefore, to accommodate CCI and to make \eqref{definition:I} correct, we set $\acir_{0}=1$. 

Following \eqref{definition:S}  and \eqref{definition:I}, we can compute the SINR for the link $(i,j)$ in RB $(f,t)$ as 
\begin{equation} \label{definition:SINR}
\gamma_{i,j,f,t} = \frac{ S_{i,j,f,t}} {\sigma^2 + (R_{j,f,t} - S_{i,j,f,t})},
\end{equation}
where $\sigma^2$ is the noise power in an RB. 

The link $(i,j)$ is said to be \textit{successful} in RB $(f,t)$ if the $\gamma_{i,j,f,t} \geq \gammaT$, which implies that the error probability is at most $\epsilon(\gammaT)$ (see Appendix~\ref{Appendix:SupportingLowErrorRate} for further details). By substituting
\eqref{definition:SINR} into $\gamma_{i,j,f,t}\ge\gammaT$ and solving for $S_{i,j,f,t}$ yields the constraint
\begin{equation}
    S_{i,j,f,t}\ge\gammabarT(\sigma^2 + R_{j,f,t}) \label{constraintSingleLink1ES},
\end{equation}
where
\begin{equation}
    \gammabarT \triangleq \frac{\gammaT}{1+\gammaT}.
\end{equation}
However, it might not be feasible to satisfy \eqref{constraintSingleLink1ES} for all links $(i,j)$ in all RBs $(f, t)$. To select which combinations of $i$, $j$, $f$, and $t$ to enforce the SINR constraint, we introduce the matrix $\mathbf{Y}\in\{0,1\}^{N\times N\times F\times T}$, where
\begin{equation}
\label{Ydefinition}
Y_{i,j,f,t} \triangleq
\begin{cases}
1, & \text{if \eqref{constraintSingleLink1ES} is enforced}\\
0, & \text{otherwise}
\end{cases}
\end{equation}

We can combine~\eqref{constraintSingleLink1ES} and~\eqref{Ydefinition} into a single constraint,
\begin{equation}
S_{i,j,f,t}\ge\gammabarT(\sigma^2 + R_{j,f,t}) - \zeta(1-Y_{i,j,f,t})		\label{constr:Yijft:compact}
\end{equation}
where $\zeta$ is a sufficiently large number to make~\eqref{constr:Yijft:compact} hold whenever $Y_{i,j,f,t}=0$, regardless of the schedule and power allocation. It is not hard to show that $\zeta = \gammabarT(\sigma^2 + NP^{\mathrm{max}})$ is sufficient. Observe that if $Y_{i,j,f,t} = 1$, then the link $(i,j)$ is successful in RB $(f,t)$ if~\eqref{constr:Yijft:compact} is satisfied. 

To make it explicit which optimization variables that affect the SINR constraint, we substitute \eqref{definition:S} and \eqref{definition:I} into \eqref{constr:Yijft:compact}, which yields
\begin{align}
P_{i,f,t}H_{i,j}\ge \gammabarT & (\sigma^2 + \sum_{f'\in\Fs}  \sum_{k\in\Ns}  P_{k,f',t}H_{k,j}\acirf)\nonumber \\
& - \zeta(1-Y_{i,j,f,t}).		\label{constr:Yijft}
\end{align}

\subsubsection{Message reception constraints}
We define the matrix $\Wb\in\{0,1\}^{N\times M\times T}$ with elements
\begin{equation}
    W_{j,m,t} \triangleq
    \begin{cases}
        1, & \parbox{6.5cm}{if message $m$ is scheduled to VUE $j$ with SINR $\geq\,\gammaT$ for the first time in timeslot $t$,}\\[0.2cm]
        0, & \text{otherwise}
    \end{cases} \label{def:Wjmt}
\end{equation}
We can compute $W_{j,m,t}$ as
\begin{equation}
    W_{j,m,t} = \Bigl(\bigvee_{i\in\Ns}   \bigvee_{f\in\Fs}  X_{i,m,f,t} \wedge Y_{i,j,f,t}\Bigr) \wedge 
    \Bigl(\neg\bigvee_{t'=0}^{t-1}  W_{j,m,t'}\Bigr)          \label{constr:Wjmt}
\end{equation}
where the \textit{AND} operation with $(\neg\bigvee_{t'=0}^{t-1} W_{j,m,t'})$ is to ensure that the message $m$ has not already been received in any of the previous timeslots $t'<t$. 

If $W_{j,m,t}=1$, then VUE $j$ can relay (i.e., transmit) the message $m$ during or after timeslot $t+t^\mathrm{p}$, where $t^\mathrm{p} \in \mathbb{Z^+}$ is the processing delay for a VUE to relay a message.  It should be noted that VUE $i$ can transmit message $m$ during timeslot $t$ if and only if (a) VUE $i$ generates the message on or before timeslot $\lfloor t-t^\mathrm{d} \rfloor$ or (b) VUE $i$ receives the message from some other VUEs on or before timeslot $t-t^\mathrm{p}$. In other words, the Boolean variable $X_{i,m,f,t}$ is constrained as\footnote{To disable multihop (i.e., relaying of messages), we replace \eqref{constr:Ximft} with the constraint $X_{i,m,f,t} \leq \bigvee_{t'=0}^t \Omega_{i,m,t'}$},
\begin{equation}
X_{i,m,f,t} \leq \Bigl(\bigvee_{t'=0}^t \Omega_{i,m,t'}\Bigr)  \vee  \Bigl(\bigvee_{t'=0}^{t-t^\mathrm{p}}   W_{i,m,t'}\Bigr).   \label{constr:Ximft}
\end{equation}


\subsubsection{Latency computation}
\label{sec:latency:computation}

The latency $\tau_{j,m}$ for the message $m$ upon reception by VUE~$j$ is computed as,
\begin{equation}
    \label{constr:latency}
    \tau_{j,m} = 
    \begin{cases}
    \sum_{t\in\Ss} t W_{j,m,t} -\arrivalTime , & \mathrm{if} \sum_{t\in\Ss} W_{j,m,t} \ge 1,\\
    \infty, &\text{otherwise.}
    \end{cases}
\end{equation}
we have adopted the convention that 
$\tau_{j,m} = \infty$
if message $m$ is never scheduled for transmission in an RB with SINR greater or equal to $\gammaT$.


\subsubsection{Age of information computation}
\label{sec:AoI:computation}
Let variable $A_{i,j,t} \in \mathbb{R^+}$ indicate the age of information of the messages from VUE $i$ to VUE $j$ at the end of timeslot $t$. With the assumption of successful reception upon satisfying the SINR threshold $\gammaT$, the variable $A_{i,j,t}$ can be computed for $t=0, 1, \ldots, T-1$ as
\begin{equation}
A_{i,j,t} = \min_{m \in \Ms_{i}} (1 + t + A^\mathrm{init}_{i,j} -  (\arrivalTime+A^\mathrm{init}_{i,j}) \sum_{t'=0}^{t} W_{j,m,t'}),   \label{constr:AOIijt}
\end{equation}
where the parameter $A^\mathrm{init}_{i,j}$ is the initial \gls{AOI} for before the start of the scheduling interval, i.e., at the beginning of timeslot $t=0$. The above equation can be translated into a set of linear constraints using the method explained in Appendix \ref{subsec:minTranslation}. 

\subsection{Latency and AoI Requirements}

It should be noted that $\tau_{j,m}$ and $A_{i,j,t}$ is the latency and AoI, respectively, if the messages that are delivered error-free to VUE~$j$ and the corresponding delivery times are exactly those indicated by $W_{j,m,t}$. In practice, however, there will be random message errors and the actual latency $\tau_{j,m}^\mathrm{E}$ and actual AoI $A_{i,j,t}^\mathrm{E}$ will therefore be random. 
Hence, it is meaningful to formulate probabilistic requirements on the latency and AoI.

The probablistic latency requirement can be formulated as
\begin{equation}
    \label{eq:latency:prob:req}
    \Pr\{\tau_{j,m}^\mathrm{E} \le \tau^\mathrm{T}\} \ge P_{\tau}^{\text{req}}
\end{equation}
where $\tau^\mathrm{T}$ is the maximum allowed latency (also known as the deadline) and $P_{\tau}^{\text{req}}$ is the required probability. The probabilistic requirement \eqref{eq:latency:prob:req} is guaranteed to be satisfied if (a) message $m$ is scheduled to arrive with latency less or equal to $\tau^{\mathrm{T}}$ and (b) the end-to-end error probability $\epsilon^{\text{e2e}}$ for the transmission of message $m$ is small enough: $(1-\epsilon^{\text{e2e}})\ge P_{\tau}^{\text{req}}$. 

As shown in Appendix~\ref{Appendix:SupportingLowErrorRate}, we adjust the SINR threshold $\gammaT$ such that the end-to-end error probability $\epsilon^{\text{e2e}}$ for all scheduled end-to-end connections is upper bounded by a given requirement $\epsilon^{\text{req}}$. To satisfy the probabilistic latency requirement, we use $\epsilon^{\text{req}}=1-P_\tau^\mathrm{req}$.

To summarize, the probabilistic latency requirement \eqref{eq:latency:prob:req} is satisfied if
\begin{subequations}
\begin{align}
    \tau_{j,m} &\le \tau^\mathrm{T}\label{eq:latency:deter:req}\\
    \epsilon^{\text{req}} &= 1-P_{\tau}^{\text{req}}\label{eq:latency:epsilon:req}.
\end{align}
\end{subequations}


Similarly, the AoI requirement can be formulated as
\begin{equation}
    \label{eq:AoI:prob:req}
    \Pr\{\mu(A_{i,j,t}^\mathrm{E}) \le \mu^\mathrm{T}\} \ge P_{A}^{\text{req}}
\end{equation}
where the metric $\mu$ is a mapping from $(A_{i,j,t}^\mathrm{E}: t\in\Ss)$ to $\mathbb{R}$, $\mu^\mathrm{T}$ is the metric threshold, and $P_{A}^{\text{req}}$ is the required probability. Without any essential loss of generality, we will limit our attention to metrics $\mu$ such that if $A_{i,j,t}' \le A_{i,j,t}$, $\forall\,t\in\Ss$, then $\mu(A_{i,j,t}') \le \mu(A_{i,j,t})$. Examples of such metrics is the time average 
\begin{equation}
    \mu(A_{i,j,t}) = \frac{1}{T} \sum_{t\in\Ss} A_{i,j,t}
\end{equation}
and time maximum
\begin{equation}
    \mu(A_{i,j,t}) = \max_{t\in\Ss} A_{i,j,t}.
\end{equation}
As shown in Appendix~\ref{appendix:AoI}, the probabilistic AoI requirement
\eqref{eq:AoI:prob:req} is satisfied if 
\begin{subequations}
\begin{align}
    \mu(A_{i,j,t}) &\le \mu^\mathrm{T}
    \label{eq:AoI:deter:req}\\
    \Rightarrow \qquad\epsilon^{\text{req}}
    &\leq
    1- (P_{A}^{\text{req}})^{1/|\Ms_i|}
    \label{eq:AoI:epsilon:req}
\end{align}
\end{subequations}
where, as usual, $\epsilon^{\text{req}}$ determines $\gammaT$ ( see~\eqref{eq:gammaT:from:epsilon}).

To conclude, we have shown how to translate probabilistic requirements on latency and AoI into the corresponding deterministic requirements augmented with appropriate requirements on the end-to-end error probability. In the following, we can therefore propose and study RRM algorithms that aim to satisfy deterministic requirements.

\subsection{Basic Problem Formulations}  \label{subsec:ProblemFormulations}

In this section, we will formulate the scheduling and power control problem as \gls{MBLP} problems for various objectives. The output of the optimization problems is therefore the schedule and power allocation matrices, $\Xb^\star$ and $\Pb^\star$, that optimize the objective function under the specified constraints. Input to the optimization problems is the slow CSI $H_{i, j}$, the set of VUE $\Ns$, the intended receiver set $\Rs_i$ for each VUE $i$, the message generation indicator $\Omega_{i,m,t}$, the ACIR function $\acir_r$, the max power constraint $\Pmax$, and the SINR threshold $\gammaT$. We recall that these variables are needed for $i\in\Ns$, $j\in\Ns$, $m\in\Ms$, $t\in\Ss$, and $r\in\Fs$.    


We recall that $W_{j,m,t}=1$ implies that message $m$ is scheduled to arrive at VUE~$j$ for the first time in timeslot $t$. The message will actually be delivered with a probability of at least $(1-\epsilon^{\text{req}})$. Hence, 
\begin{equation}
    (1-\epsilon^{\text{req}})
    \sum\limits_{m \in \Ms_{i}}\sum\limits_{t\in\Ss}  W_{j,m,t}
\end{equation}
is a lower bound on the throughput
(i.e., the expected number of unique delivered messages in $T$ timeslots)
from VUE~$i$ to VUE~$j$ . With a slight abuse of terminology, we will call $\sum_{m \in \Ms_{i}}\sum_{t\in\Ss}  W_{j,m,t}$ “throughput” in problem formulations 1) and 2) below.

\subsubsection{Maximizing throughput}
The problem to maximize the total sum-throughput of the network can be formulated as
\begin{subequations} \label{formulation:maximizeThroughput}
	\begin{align}
	&\max\limits_{\mathbf{P}, \mathbf{X}, \mathbf{Y}, \mathbf{W}}      \sum\limits_{i \in \Ns}  \sum\limits_{j \in \Rs_i}  \sum\limits_{m \in \Ms_{i}}\sum\limits_{t\in\Ss}  W_{j,m,t}  \\			
	& \mbox{subject to,}    \nonumber  \\
	& 
	\eqref{constr:Xift}, 
	\eqref{constr:power_total}, 
	\eqref{constr:PConstrainedByX},  
	\eqref{constr:Yijft},   
	\eqref{constr:Wjmt},  
	\eqref{constr:Ximft}   \nonumber
	\end{align}
\end{subequations}

\subsubsection{Maximizing the worst-case throughput}
The problem to maximize the minimum throughput of an end-to-end connection in the network can be formulated as
\begin{subequations} \label{formulation:maximizeWorstThroughput}
	\begin{align}
	&\max_{\mathbf{P}, \mathbf{X}, \mathbf{Y}, \mathbf{W}}   \eta^\mathrm{min}  \\			
	& \mbox{subject to,}    \nonumber  \\
	& \sum\limits_{m \in \Ms_{i}}\sum\limits_{t\in\Ss}  W_{j,m,t} \geq \eta^\mathrm{min},\qquad i\in\Ns,j \in \Rs_i\\
	& 
	\eqref{constr:Xift}, 
	\eqref{constr:power_total}, 
	\eqref{constr:PConstrainedByX},  
	\eqref{constr:Yijft},   
	\eqref{constr:Wjmt},  
	\eqref{constr:Ximft}   \nonumber
	\end{align}
\end{subequations}
\subsubsection{Maximizing the connectivity}
\label{subsec:ProblemFormulations_MaxConnectivity}
VUE $i$ and VUE $j$ are said to be connected if at least one message can be sent from $i$ to $j$ with the required end-to-end error probability during the scheduling interval. Let $Z_{i,j}\in\{0, 1\}$ indicate that VUE $i$ and $j$ are connected, then 
\begin{equation}
    Z_{i,j} = 
    \min\Bigl\{1, 
    \sum_{j \in \Rs_i}
    \sum_{m \in \Ms_{i}}  \sum_{t\in\Ss}  W_{j,m,t}
    \Bigr\}
\end{equation}
Hence, the following problem maximizes the network connectivity,
\begin{subequations} \label{formulation:maximizeConnectivity}
	\begin{align}
	&\max_{
	\mathbf{W}, \mathbf{X}, \mathbf{Y}, \mathbf{P}, \mathbf{Z}}   
	\sum_{i \in \Ns}  
	\sum_{j \in \Rs_i}  Z_{i, j}\\			
	& \mbox{subject to,}    \nonumber  \\
	&Z_{i, j} \le \sum_{t=0}^{T-1} \sum_{j \in \Rs_i} \sum_{m \in \Ms_{i}}  W_{j,m,t},\qquad \forall i, j\in\Rs_i \label{constr:Zij} \\
	& Z_{i,j} \leq 1 \\
    & \eqref{constr:Xift}, \eqref{constr:power_total}, \eqref{constr:PConstrainedByX},  \eqref{definition:S},  \eqref{definition:I},   \eqref{constr:Yijft},   \eqref{constr:Wjmt},  \eqref{constr:Ximft}   \nonumber
	\end{align}
\end{subequations}

Note that, we can formulate the problem to maximize the minimum connectivity for a VUE in the network, i.e.,   $\max(\min_{i\in\Ns}\sum_{j\in\Rs_i} Z_{i, j})$, by a similar transformation as in~\eqref{formulation:maximizeWorstThroughput}.

\subsubsection{Maximizing connectivity for AoI requirements}

Suppose $\gammaT$ is chosen such that~\eqref{eq:AoI:epsilon:req} is satisfied, then the following problem will maximize the number of end-node pairs $(i, j)$ such that the probabilistic AoI requirement~\eqref{eq:AoI:prob:req} is satisfied:
\begin{subequations} \label{formulation:maximizeConnectivityAOI}
	\begin{align}
	&\max_{\mathbf{P}, \mathbf{X}, \mathbf{Y}, \mathbf{W}, \mathbf{Z}}
	\sum_{i \in \Ns}  
	\sum_{j \in \Rs_i}  Z_{i,j}^\mathrm{A} \\			
	& \mbox{subject to,}    \nonumber  \\
	& \mu(A_{i,j,t}) \leq \mu^\mathrm{T} +  \zeta (1-Z_{i,j}^A), \qquad i\in\Ns, j \in \Rs_i\label{constr:AOI:deter:req} \\
	& Z_{i,j}^A \in \{0,1\} \quad \forall\, i,j\\
	& 
	\eqref{constr:Xift}, 
	\eqref{constr:power_total}, 
	\eqref{constr:PConstrainedByX},  
	\eqref{constr:Yijft},   
	\eqref{constr:Wjmt},  
	\eqref{constr:Ximft},  
	\eqref{constr:AOIijt}    
	\nonumber
	\end{align}
\end{subequations}
where $Z_{i,j}^A = \mathbbm{1}\{\mu(A_{i,j,t}) \leq \mu^\mathrm{T}\}$ indicates if the deterministic \gls{AOI} requirement is satisfied for the end-node pair $(i,j)$ and $\zeta$ is chosen large enough such that \eqref{constr:AOI:deter:req} holds when $Z_{i,j}^A=0$. Clearly, if $\mu(A_{i,j,t})\le\mu^{\max}$, then $\zeta = \mu^{\max}-\mu^\mathrm{T}$ is sufficient. For instance, if $\mu(A_{i,j,t}) = \max_{t\in\Ss} A_{i,j,t}$, then $\zeta = T + \max_{i\in\Ns, j\in\Rs_j} A_{i,j}^{\text{init}} -\mu^\mathrm{T}$ is sufficiently large.

Problem \eqref{formulation:maximizeConnectivityAOI} is an \gls{MBLP} if \eqref{constr:AOI:deter:req} can be translated into a number of linear constraints. This is possible if $\mu$ is a linear or one-to-one mapping, or if $\mu(A_{i,j,t}) = \max_{t\in\Ss} A_{i,j,t}$.

\subsubsection{Maximizing connectivity for latency requirement}

To formulate a problem that maximizes the number of end-node pairs $(i, j)$ such that the probabilistic latency requirement~\eqref{eq:latency:prob:req} is satisfied follows the same logic as for AoI. Suppose $\gammaT$ is chosen such that~\eqref{eq:latency:epsilon:req} is satisfied, then the following problem will achieve the end goal:
\begin{subequations} \label{formulation:maximizeConnectivity:latency}
	\begin{align}
	&\max_{\mathbf{P}, \mathbf{X}, \mathbf{Y}, \mathbf{W}, \mathbf{Z}}
	\sum_{i \in \Ns}  
	\sum_{j \in \Rs_i}  Z_{i, j}^\tau \\			
	& \mbox{subject to,}    \nonumber  \\
	& \tau_{j,m} \leq \tau^{\mathrm{T}} +  \zeta (1-Z_{i,j}^\tau), \qquad i\in\Ns, j \in \Rs_i
	\label{constr:latency:deter:req} \\
	& Z_{i,j}^\tau \in \{0,1\} \quad \forall\, i,j\\
	& 
	\eqref{constr:Xift}, 
	\eqref{constr:power_total}, 
	\eqref{constr:PConstrainedByX},  
	\eqref{constr:Yijft},   
	\eqref{constr:Wjmt},  
	\eqref{constr:Ximft},  
	\eqref{constr:latency}    
	\nonumber
	\end{align}
\end{subequations}
where $Z_{i,j}^\tau = \mathbbm{1}\{\tau_{j,m} \leq \tau^\mathrm{T}\}$ indicates if the deterministic latency requirement is satisfied for the end-node pair $(i,j)$ and $\zeta$ is chosen large enough such that \eqref{constr:latency:deter:req} is holds when $Z_{i,j}^\tau = 0$. A technical problem arise here, since, by convention, $\tau_{j,m}=\infty$ if message $m$ is not scheduled to be transmitted in an RB where the SINR at VUE~$j$ is larger or equal to $\gammaT$. We can resolve this technicality in~\eqref{constr:latency} by replacing $\infty$ by a very large number, say $10^{10}$, and set $\zeta$ to the same number. (In a practical implementation with finite-precision arithmetic, more reasonable numbers must, of course, be used.)

\subsection{Variations of Basic Problem Formulations}

We note that all of the problem formulations in Section~\ref{subsec:ProblemFormulations} can be made to power control alone problem by fixing $X_{i,m,f,t}$ and to scheduling alone problem by modifying \eqref{constr:PConstrainedByX} to $P_{i,f,t} = \bar{P}_{i,t}\tilde{X}_{i,f,t}$, where $\bar{P}_{i, t}$ is the transmit power of VUE $i$ if scheduled in timeslot $t$. Under the assumption that all VUEs use the same transmit power, $\bar P_{i,t}=\bar P_{t}$ for all $i$, then $\bar{P}_{t} = \Pmax$ maximizes the performance for all scheduling alone algorithms, as proved in \cite{Anver_Access1}. The resulting problems are \gls{MBLP} and \gls{BLP} problems, respectively.

Moreover, scheduling and power control can be done for certain subset of timeslots alone. Assume that we are interested in scheduling and power control on and after timeslot $T'<T$ only, and we know the packet reception status for all the timeslots prior to timeslot $T'$. Then we can set all variables (i.e., $\mathbf{P},\mathbf{X},\mathbf{Y},\mathbf{W}$) corresponding to timeslots $\{0,1,\dots,T'-1\}$, and optimize over all variables corresponding to timeslots $\{T',T'+1,\dots,T\}$. This way, we can accommodate the information upon any past transmissions and \gls{AOI}. For instance, the scheduling and power control can be done for all timeslots one by one, to reduce computational complexity. Other practical considerations such as supporting large message payload and high reliability, are discussed in Appendix \ref{Appendix:PracticalConsiderations}.

\section{Clustering of Network}  \label{sec:Clustering}
The basic problem formulations in Section~\ref{subsec:ProblemFormulations} have computational complexities that scale poorly with the network size $N$. To address this issue, we propose to partition the network into smaller groups and perform resource allocation in each group independently. 

The entire network is partitioned into $C$ clusters and each cluster into $G$ groups. A group $g$ in cluster $c$ is called group $(c,g)$. The notion of a group is similar to a cell in a traditional cellular system. The available time slots will be partitioned over the groups belonging to a cluster, which will eliminate intergroup interference since groups $(c,g)$ and $(c,g')$,  will use nonoverlapping timeslots when $g'\neq g$. We can therefore do scheduling and power control for each group in a cluster independently, which greatly reduces the computational complexity as the network size $N$ increases. 

However, since groups $(c, g)$ and $(c', g)$ can use the same timeslots, there will in general be intercluster interference. It is therefore important to design the clusters and groups such that this interference is controlled. In the following, we will propose a method for this. 

We start by defining some notation.
Let $\Ts^{(c,g)} \subseteq \Ns$ denote the set of Tx-VUEs in group $(c,g)$ for $c=0, 1, \ldots, C-1$ and $g=0, 1, \ldots, G-1$. Moreover, let 
\begin{equation}
    \Rs^{(c,g)} = \{j: i \in \Ts^{(c,g)}, \, j\in \Rs_i \}  \label{compute:Rcg}  
\end{equation}
be the set of Rx-VUEs intended to receive messages from VUEs in group $(c,g)$. 

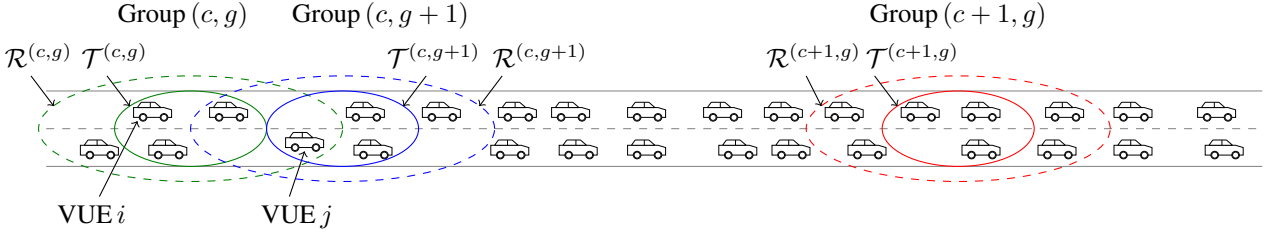
\begin{figure*}[t]	 
	\begin{flushleft}
		\newcommand{\scaleValueOfCar}{0.1}
\begin{tikzpicture}[
carDrawing/.pic={ 
	\draw[scale=\scaleValueOfCar,color=black,very thin] (1.5,.5) -- ++(0,1.3) -- ++(1,0) --  ++(3,0) -- ++(1,-0.3) -- ++(0,-1) -- (1.5,.5) -- cycle;    	
	\draw[scale=\scaleValueOfCar,rounded corners=0.05ex,fill=white,very thin]  (2,1.8) -- ++(0.5,0.7) -- ++(1.8,0) -- ++(0.8,-0.7) -- cycle;			  
	\draw[scale=\scaleValueOfCar,very thin]  (3.7,1.8) -- (3.7,2.5);
	\draw[scale=\scaleValueOfCar,draw=black,fill=white!50,very thin] (2.75,.5) circle (.5);	
	\draw[scale=\scaleValueOfCar,draw=black,fill=white!50,very thin] (5.5,.5) circle (.5);
	\draw[scale=\scaleValueOfCar,draw=black,fill=white!80,very thin] (2.75,.5) circle (.4);
	\draw[scale=\scaleValueOfCar,draw=black,fill=white!80,very thin] (5.5,.5) circle (.4);
}]

\draw[line width=0.1,color=black!50] (0,0.5cm)--(16cm,0.5cm);
\draw[line width=0.1,dashed,color=black!50] (0,0)--(16cm,0);
\draw[line width=0.1,color=black!50] (0,-0.5cm)--(16cm,-0.5cm);


\draw (1cm,0.1cm) node(VUEi){} pic[] {carDrawing};   
\draw (2cm,0.1cm) pic[] {carDrawing};  
\draw (3.8cm,0.1cm) node(VUEj){} pic[] {carDrawing};   
\draw (4.8cm,0.1cm) pic[] {carDrawing}; 
\draw (5.8cm,0.1cm) pic[] {carDrawing};   
\draw (6.5cm,0.1cm) pic[] {carDrawing}; 
\draw (7.5cm,0.1cm) pic[] {carDrawing};   

\draw (8.5cm,0.1cm) pic[] {carDrawing};   
\draw (9.3cm,0.1cm) pic[] {carDrawing}; 
\draw (10.1cm,0.1cm) pic[] {carDrawing};   
\draw (11.1cm,0.1cm) pic[] {carDrawing}; 
\draw (11.9cm,0.1cm) pic[] {carDrawing};   
\draw (13cm,0.1cm) pic[] {carDrawing};
\draw (13.9cm,0.1cm) pic[] {carDrawing};   
\draw (15cm,0.1cm) pic[] {carDrawing};   

\draw (0.3cm,-0.4cm) pic[] {carDrawing};   
\draw (1.2cm,-0.4cm) pic[] {carDrawing};   
\draw (3cm,-0.3cm) node(VUEkA){} pic[] {carDrawing};   
\draw (3.9cm,-0.4cm) pic[] {carDrawing};   
\draw (5.7cm,-0.4cm) pic[] {carDrawing};   
\draw (6.6cm,-0.4cm) pic[] {carDrawing};   
\draw (7.5cm,-0.4cm) pic[] {carDrawing};   

\draw (8.7cm,-0.4cm) pic[] {carDrawing};   
\draw (9.4cm,-0.4cm) pic[] {carDrawing}; 
\draw (10.4cm,-0.4cm) pic[] {carDrawing};   
\draw (11.9cm,-0.4cm) pic[] {carDrawing}; 
\draw (12.9cm,-0.4cm) pic[] {carDrawing};   
\draw (13.9cm,-0.4cm) pic[] {carDrawing};
\draw (15.1cm,-0.4cm) pic[] {carDrawing};

\draw[->,black]  ($(VUEi)+(-0.3cm,-1cm)$) node[black,below,xshift=-0.1cm]{VUE\,$i$} -- ($(VUEi)+(0.2cm,0cm)$);
\draw[->,black]  ($(VUEkA)+(0.2cm,-0.6cm)$) node[black,below,xshift=0.1cm]{VUE\,$j$} -- ($(VUEkA)+(0.4cm,0cm)$);

\coordinate (cluster1center) at (1.9cm,0cm);
\node[above,xshift=-0.1cm,yshift=1.2cm] at (cluster1center) {Group\,$(c,g)$};
\draw[color=green!50!black]  (cluster1center) ellipse (1cm and 0.5cm);
\draw[color=green!50!black,dashed]  (cluster1center) ellipse (2cm and 0.7cm);

\draw[->] ($(cluster1center)+(-1.1cm,0.7cm)$) node[above,xshift=0.1cm]{$\Ts^{(c,g)}$} -- ($(cluster1center)+(-0.8cm,0.3cm)$);
\draw[->] ($(cluster1center)+(-2.1cm,0.7cm)$) node[above,xshift=0.1cm]{$\Rs^{(c,g)}$} -- ($(cluster1center)+(-1.8cm,0.3cm)$);

\coordinate (cluster1center) at (3.9cm,0cm);
\node[above,xshift=0.5cm,yshift=1.2cm] at (cluster1center) {Group\,$(c,g+1)$};
\draw[color=blue]  (cluster1center) ellipse (1cm and 0.5cm);
\draw[color=blue,dashed]  (cluster1center) ellipse (2cm and 0.7cm);

\draw[->] ($(cluster1center)+(1.1cm,0.7cm)$) node[above,xshift=0.1cm]{$\Ts^{(c,g+1)}$} -- ($(cluster1center)+(0.8cm,0.3cm)$);
\draw[->] ($(cluster1center)+(2.1cm,0.7cm)$) node[above,xshift=0.5cm]{$\Rs^{(c,g+1)}$} -- ($(cluster1center)+(1.8cm,0.3cm)$);

\coordinate (cluster2center) at (12cm,0cm);
\node[above,yshift=1.2cm] at (cluster2center) {Group\,$(c+1,g)$};
\draw[color=red] (cluster2center) ellipse (1cm and 0.5cm);
\draw[color=red,dashed]  (cluster2center) ellipse (2cm and 0.7cm);

\draw[->] ($(cluster2center)+(-1.1cm,0.7cm)$) node[above,xshift=0.5cm]{$\Ts^{(c+1,g)}$} -- ($(cluster2center)+(-0.8cm,0.3cm)$);
\draw[->] ($(cluster2center)+(-2cm,0.7cm)$) node[above,xshift=0.1cm]{$\Rs^{(c+1,g)}$} -- ($(cluster2center)+(-1.8cm,0.3cm)$);

\end{tikzpicture} 
	\end{flushleft}
	\centering	
	\caption{Groups $(c,g)$ and $(c+1,g)$ are separated by more than the reuse distance and can therefore reuse timeslots.}\label{drawing:carsOnStreet}
			\label{Fig:carsOnStreet}
\end{figure*}

We note that the transmitter groups form a partitioning of the network, i.e., $\Ts^{(c,g)}  \cap \Ts^{(c',g')} = \emptyset$ for $(c, g)\neq (c', g')$ and the union of all transmitter groups contains all $N$ VUEs. However, the Rx-VUE sets $\Rs^{(c,g)}$ can be overlapping, see Fig.~\ref{Fig:carsOnStreet}.

\subsection{1-Hop Feasibility}

Consider the link $(i, j$) between VUE~$i$ and VUE $j$. If the channel gain $H_{i, j}$ is large enough to allow for direct (one-hop) communication with SNR greater or equal to $\gammaT$, then we say that the link $(i,j)$ is \emph{1-hop feasible}. If the link is not 1-hop feasible, we must rely on multihop communication (relaying) to connect VUE~$i$ and $j$. 

Suppose VUE~$i$ is transmitting on RB $(f,t)$. Then the SINR for the link $(i, j)$ can be upper-bounded as
\begin{align}
    \frac{S_{i,j,f,t}}{\sigma^2 + R_{j,f,t} - S_{i,j,f,t}} 
    \le \frac{S_{i,j,f,t}}{\sigma^2}
    \le \frac{\Pmax H_{i, j}}{\sigma^2}.
\end{align}
Clearly, the link $(i, j)$ is 1-hop feasible only if
\begin{equation}
    H_{i,j} \ge \gammaT\sigma^2/\Pmax,
\end{equation}
and we define the set of receivers to VUE~$i$ that are 1-hop feasible as
\begin{equation}
    \label{eq:D:def}
    \mathcal{D}_i \triangleq \{j\in\Ns: H_{i,j} \ge \gammaT\sigma^2/\Pmax\}.
\end{equation}
Note that we are considering all VUEs as potential receivers in the definition of $\mathcal{D}_i$, not only the  receivers in $\Rs_i$, since we want to consider also the case when VUE~$i$ relays messages to VUEs that are not in $\Rs_i$ (its set of intended end receivers). We note that the set of VUEs that can be reached by a VUE in $\Ts^{(c,g)}$ is
\begin{equation}
    \mathcal{D}^{(c,g)} = \bigcup_{i\in\Ts^{(c,g)}}\mathcal{D}_i.
\end{equation}

\subsection{Reuse Distance and Clustering}  \label{sec:ClusteringBasicIdea}
When clustering, we want to limit the intercluster interference to the receivers in $\mathcal{D}_i$ for all $i$.
There will be no intergroup interference, since distinct groups will use nonoverlapping timeslots. However, timeslots are reused among clusters, i.e., transmitters in the same group but different clusters can use the same timeslot. Suppose $i\in\Ts^{(c,g)}$, then the intercluster interference to receiver $j\in\mathcal{D}_i$ is
\begin{align}
    \sum_{\substack{k\in\Ts^{(c',g)}\\ c'\neq c}}&\sum_{f'=0}^{F-1} \tilde{X}_{k,j,f',t} P_{k, j, f', t} H_{k,j}\lambda_{|f'-f|}\nonumber\\
    &\approx \sum_{\substack{k\in\Ts^{(c',g)}\\ c'\neq c}}\Pmax H_{k,j}\\
    &\approx 2\Pmax \max_{\substack{k\in\Ts^{(c',g)}\\ c'\neq c}} H_{k,j}\label{eq:CCI:bound}
\end{align}
where in the first approximation, we have assumed the worst-case CCI (max power and CCI from all clusters) and ignored the ACI, and in the second approximation, we have ignored all CCI terms except for the two largest terms and replaced the second largest term with the largest one (resulting in the factor 2). The motivation for the first approximation is that ACI is negligible compared to CCI. The motivation for the last approximation is that we have at most one CCI interferer per cluster and that in a highway scenario, we have at most two neighboring clusters to cluster $c$. Hence, in the last approximation, we ignore CCI from non-neighboring clusters and upper bound the CCI from the neighboring clusters.

\begin{algorithm}[t]   
	\caption{Clustering Algorithm}  \label{Alg:Clustering}
	\begin{algorithmic}[1]
		\Require{$\{N,\Hb,\delta,\sigma^2,\Pmax,\nTx\}$}
		\Ensure{$C,G,\Ts^{(c,g)},\Rs^{(c,g)}$, $\Ss_g$}
		\State Compute $G$ using  \eqref{compute:d_reuse} and \eqref{compute:G}
		\State Compute $C$ using~\eqref{compute:C}.
		\State Compute $\Ts^{(c,g)}$ and $\Rs^{(c,g)}$, $\forall(c, g)$ using \eqref{compute:Tcg} and \eqref{compute:Rcg}
		\State Compute $\Ss_g$, $\forall g$ using~\eqref{compute:Sg} 
	\end{algorithmic}
\end{algorithm}

When clustering, we strive to set the reuse distance sufficiently large such that the intercluster interference does not exceed $\delta \sigma^2$ from some (small) $\delta$. Using the approximation \eqref{eq:CCI:bound}, we can achieve this if
\begin{equation}
    H_{k,j}
    \le 
    \frac{\delta\sigma^2}{2\Pmax}, \quad i\in\Ts^{(c, g)}, j\in\mathcal{D}_i, k\in\Ts^{(c',g)}, c'\neq c, \forall g.
\end{equation}
We ensure this by setting $|k-i|> d^{\text{reuse}}$ for all VUEs $i$ and $k$ that belong to the same group but different clusters, where
\begin{equation}
    d^{\text{reuse}} = \max_{i\in\Ns, k\in\Ns}\bigl\{|k-i|: H_{k, j} > \frac{\delta\sigma^2}{2\Pmax}, j\in\mathcal{D}_i\bigr\}.  \label{compute:d_reuse}
\end{equation}
Given that each group should include $\nTx$ VUEs, we can compute the number of groups $G$ and the number of clusters $C$ as
\begin{align}
    G &= \left\lceil\frac{\nTx+d^{\text{reuse}}}{\nTx}\right\rceil   \label{compute:G} \\
    C &= \left\lceil\frac{N}{G\nTx}\right\rceil \label{compute:C}
\end{align}
We can now form the group $(c,g)$ for $c=0, 1, \ldots, C-1$ and $g=0, 1, \ldots, G-1$ as
\begin{equation}
    \Ts^{(c,g)} = \{(cG+g) \nTx + n: n = 0, 1, \ldots, \nTx-1\},
    \label{compute:Tcg}
\end{equation}
and $\Rs^{(c, g)}$ follows from~\eqref{compute:Rcg}. We will assign the the timeslots in $\Ss_g$ to group $(c,g)$, where
\begin{equation}
    \Ss_g \triangleq
    \{g+\ell G: g+\ell G <T, 0 \leq \ell \leq \lfloor T/G \rfloor\}
    \label{compute:Sg}
\end{equation}
The clustering procedure is summarized in Algorithm~\ref{Alg:Clustering}.

\subsection{Scheduling and Power Control} \label{sec:CCSP}
As mentioned above, we can perform scheduling and power control for each group independently. The required computations can be done in a completely centralized fashion by a single controller or be  distributed to at most $CG$ controllers, one per group in the network. Hybrids of these extreme architectures are, of course, also possible. The controllers can be hosted by fixed infrastructure nodes (e.g., in BSs or edge computing devices) or by specially assigned VUEs. 

All problem formulations in Section~\ref{subsec:ProblemFormulations} can used to compute the schedule and power control for a particular group after the following modifications. 

Firstly, we need to modify the SINR constraint~\eqref{constr:Yijft} to add a margin for the inter-cluster interference,
\begin{equation}
S_{i,j,f,t}\ge\gammabarT(\sigma^2(1+\delta) + R_{j,f,t}) - \zeta(1-Y_{i,j,f,t}).		\label{constr:Yijft:clustered}
\end{equation}

Secondly, we reduce the matrices $\mathbf{P}$, $\mathbf{X}$, $\mathbf{Y}$, $\mathbf{W}$, and  $\mathbf{Z}$ and limit the input variables $H_{i,j}$ and $\Omega_{i, m, t}$ to cover only the relevant variables for the group. In general, we need to consider only the transmitters $i\in\Ts^{(c,g)}$, receivers  $j\in\Rs^{(c,g)}$, messages $m\in\Ms^{(c,g)} = \cup_{i\in\Ts^{(c,g)}}\Ms_i$, and timeslots $t\in\mathcal{S}^g$. 

In particular, this will reduce the dimensions of the matrices such that
\begin{align}
\mathbf{P}&\in\{0, 1\}^{\nTx\times F\times T_g}, \label{compute:PforaGroup}\\ 
\mathbf{X}&\in\{0, 1\}^{\nTx\times M \times F \times T_g},\\ 
\mathbf{Y}&\in\{0, 1\}^{\nTx\times \nRx_g \times F \times T_g},\\ 
\mathbf{W}&\in\{0, 1\}^{\nRx_g\times M_g \times T_g}, \\  
\mathbf{Z}&\in\{0, 1\}^{\nRx_g\times \nRx_g},  \label{compute:ZforaGroup}
\end{align}
where $T_g = |\Ss_g|$, $\nRx_g = |\Rs^{(c,g)}|$, and $M_g = |\Ms^{(c,g)}|$.


The main advantage with clustering is a reduction of the overall computational complexity (see Section~\ref{sec:ComputationalComplexity}). The main drawbacks is (a) a potential loss of performance (since clustering cannot improve the optimal values of objective functions) and (b) that multihop communication (relaying) is only possible between end-nodes $i\in\Ts^{(c,g)}$ and $j\in\mathcal{D}^{(c,g)}$. Hence, any intended receiver $j\in\Rs^{(c,g)}\setminus\mathcal{D}^{(c,g)}$ cannot receive any messages. 

Within the current framework, such connections can only be enabled by relaying through the fixed infrastructure, i.e., the messsage is transmitted from the source VUE~$i$ via an uplink to its serving \gls{BS}, which in turn forwards the message to a base station that can reach the destination VUE~$j$ via a downlink. This will, require some (minor) modifications of the basic framework in this paper, incur extra latency that might not be acceptable, and will only work when VUEs are inside coverage of the fixed network infrastructure. Finding a better relaying strategy for long connections is, however, outside the scope of this paper. We also note that this problem is most prevalent when the source and destination VUEs are far from each other and the source at the edge of the group. Hence, the problem might not be so serious, since most traffic safety applications with low latency requirements rely on communication over relative short distances.

\section{Clustering-Based Distributed Scheduling (CDS)}  \label{sec:CDS}

\begin{algorithm}[t]
	\caption{CDS}      \label{Alg:CDS}
	\begin{algorithmic}[1]
		\Require{$\{i',\nTx,T, F,G,\beta,\lambda\}$}
		\Ensure{$\tilde{\Xb}$}
		\State $\mathbf{\tilde{X}} = \textbf{0}^{\nTx \times F \times T}$
		\State $\Js = \Ts^{(c,g)}$ // unscheduled VUEs  
		\State $\Us = \{(f, t): f\in\Fs, t\in\Ss_g \}$ // unscheduled RBs
		\State $c=\lfloor i'/(GN^\mathrm{Tx}) \rfloor$
		\State $g = \lfloor i'/\nTx \rfloor \bmod  G$
		\State Compute $\Ts^{(c,g)}$ from $c$, $g$, and $\nTx$ using \eqref{compute:Tcg} 
		\State Compute $\Ss_g$ from $g$, $T$, and $G$ using \eqref{compute:Sg}
		\State //  Stage 1: Schedule all VUEs in group $(c,g)$ exactly once
		\Do
		\State $(i^*, f^*, t^*) = \argmin\limits_{
		    \{(i, f, t): i\in\Js, (f,t)\in\Us\}}
		    \sigma_I^2(i, f, t; \tilde{\Xb})$
		\State $\tilde{X}_{i^*,f^*,t^*} = 1$
		\State $\Js = \Js \setminus \{i^*\}$
		\State $\Us = \Us \setminus \{(f^*,t^*)\}$
		\doWhile $\Js \neq \emptyset $
		\State //  Stage 2: Allocate remaining unscheduled RBs
		\For{$(f,t)\in\Us$}
		\State $i^* = \argmin\limits_{i\in\Ts^{(c,g)}:\vee_{f'=0}^{F-1} \tilde{X}_{i,f',t}=0}
		\sigma_I^2(i, f, t; \tilde{\Xb})$
		\State $\tilde{X}_{i^*,f,t} = 1$		 
		\EndFor
	\end{algorithmic}
\end{algorithm}

The RRM solutions described above are centralized in the sense that the computation of the schedule and power allocation is performed by a central controller (e.g., a \gls{BS} or a specially elected VUE). In this section, we will present a cluster-based scheduling algorithm that is fully distributed. That is, a scheme in which each VUE computes its group schedule, independently of the other VUEs. The algorithm, called 
clustering-based distributed scheduling (CDS), requires knowledge of the  position index $i'$ of an arbitrary VUE in the group, the system parameters $T$, $F$, $\nTx$, $G$, and the ACIR function $\lambda$. The algorithm has also a tuning parameter $\beta$, which is described below. It should be noted that CDS does not require the channel state matrix $\Hb$. The output is the scheduling matrix $\Xb$, which can be used with any power matrix, although CDS is designed with the tacit assumption of equal transmit powers. 


The rationale behind the algorithm is based on the assumption that each VUE would like to transmit messages to its nearby VUEs. Hence, the interference the \emph{transmitter} VUE~$i$ experiences is approximately equal to the interference its intended \emph{receiver} VUE~$j$, $j\in\Rs_i$, experiences. Hence, it makes sense to schedule the VUE~$i$ \textit{transmission} on an RB $(f,t)$ where VUE~$i$ would experience low \textit{received} interference. Since the interference depends on the schedule, we construct the schedule in an iterative, greedy fashion.

To be more precise, we have at the beginning of an iteration access to the partial schedule $\tilde{\Xb}$ constructed so far. The intragroup interference that VUE~$i\in\Ts^{(c,g)}$ experiences in RB $(f,t)$ is
\begin{equation}
    \sum_{k\in\Ts{(c,g)}}\sum_{f'\in\Fs} 
    \bar{P} \tilde{X}_{k, f', t} H_{k,i}\acir_{|f'-f|} 
\end{equation}
where $\bar{P}$ is the common transmit power for all VUEs. Ignoring pathloss and assuming that each blocking vehicle introduce an additional gain $\beta<1$, we note that $H_{k,i} = \beta^{|k-i|-1}$. Hence, the intragroup interference is proportional to
\begin{equation}
    \sigma_I^2(i, f, t; \tilde{\Xb}) \triangleq \sum_{k\in\Ts{(c,g)}}\sum_{f'\in\Fs}  
    \tilde{X}_{k, f', t}\acir_{|f'-f|}\beta^{|k-i|-1}.
\end{equation}
The main idea is to iteratively identify the triplet $(i^*, f^*, t^*)$ that minimizes $\sigma_I^2(i, f, t; \tilde{\Xb})$ (under some suitable constraints) and schedule VUE~$i^*$ in RB $(f^*, t^*)$. The process is then repeated until a termination criterion is met.

We propose to construct the schedule in three steps. In the first step, we ensure that each VUE in the group is scheduled exactly once. In the second step, we assign any unscheduled RBs to the VUEs in the group (without attempting to keep the number of RBs assigned equal for all VUEs). In the third step, we assign messages to the scheduled RBs. 

For simplicity, we describe the algorithm for the case when all VUEs have exactly one own message to transmit during the scheduling interval and that this message is available at timeslot $t=0$. The extension to a more general data traffic model is not difficult, but would complicate the presentation here. 

The first two steps are summarized in Alg.~\ref{Alg:CDS}, which outputs $\tilde{\Xb}$. Ties in the $\argmin$ operations in Alg.~\ref{Alg:CDS}  are resolved to the smallest value of $i$, $f$, and $t$. Due to this, the VUE with the lowest index in the group will be scheduled in RB $(0, g)$, the second lowest VUE will be scheduled in $(0, g+G)$, etc. Once all timeslots in $\Ss_g$ been scheduled once, VUEs will start to be multiplexed in frequency. Note that, a VUE is scheduled at most once in a timeslot to avoid sharing of transmit power among RBs in a timeslot. 

What remains is the third step: to assign messages indices to the scheduled transmissions, i.e., convert $\tilde{X}_{i,f,t}$ to $X_{i,m,f,t}$. This can be done in many ways. At a scheduled RB $(f,t)$, the VUE can choose to transmit its own message or transmit (i.e., relay) any other message that was received at or before timeslot $t-t^\mathrm{p}$. A reasonable strategy is for each VUE to (a) transmit is own message at the earliest scheduled timeslot and (b), for any future scheduled timeslots, relay a message that was received from the furthest located VUE. Messages are only relayed once, and if no message is available for relaying, the VUE transmits its own message again. Rule (a) strives to disseminate the original messages as quickly as possible inside the group, and rule (b) strives to relay messages as far as possible for each hop. 

\section{Computational Complexity Analysis} \label{sec:ComputationalComplexity}

In general, the worst case complexity of an \gls{MBLP} problem with $m$ Boolean variables and $n$ continuous variables can be upper-bounded as $\mathcal{O}(\frac{n^3 2^m}{\log n})$. The complexity $2^m$ is for fixing $m$ Boolean variables, and the complexity $\frac{n^3}{\log n}$ is for solving each of the resulting \gls{LP} problem using an interior point method \cite{Florian1}. Since $\Xb, \Yb, \Wb$ are Boolean and $\Pb,\Zb$ are continuous variable matrices, we can upper-bound the worst-case computational complexity of all problem formulations with objectives throughput or connectivity as $\mathcal{O}(CG\frac{(|\Pb|+|\Zb|)^3 2^{|\Xb|+|\Yb|+|\Wb|})}{\log (|\Pb|+|\Zb|)})$, where operation $|\cdot|$ indicate the number of elements in the matrix, and $CG$ account for the total number of groups in the network. 
Similarly for the problem formulation \eqref{formulation:maximizeConnectivityAOI} and \eqref{formulation:maximizeConnectivity:latency} the complexity are $\mathcal{O}(CG\frac{((|\Pb|+|\Ab|)^3 2^{|\Xb|+|\Yb|+|\Wb|+|\Zb^\mathrm{A}|})}{\log (|\Pb|+|\Ab|)})$ and $\mathcal{O}(CG\frac{((|\Pb|+|\Ab|)^3 2^{|\Xb|+|\Yb|+|\Wb|+|\Zb^{\tau}|})}{\log (|\Pb|+|\Ab|)})$. Note that the significant computational complexity reduction due to clustering is mainly due to the reduction of the size of matrices as shown in \eqref{compute:PforaGroup}--\eqref{compute:ZforaGroup}.

For Algorithm \ref{Alg:CDS}, there are $\nTx$ iterations in the first stage, and $\max\{0,FT_g-\nTx\}$ iterations in the second stage. Within an iteration in the first stage, the algorithm has to search through all $\nTx FT_g$ possible combinations of scheduling, hence, complexity is $\mathcal{O}((\nTx)^2 FT_g)$. Similarly, the computational complexity of second stage is at most $\mathcal{O}(\nTx FT_g)$. A VUE can transmit maximum $FT_g$ messages, therefore, the worst-case complexity for stage 3 is $\mathcal{O}(\nTx FT_g)$. In summary, the worst-case complexity of Algorithm \ref{Alg:CDS} is upper-bounded by $\mathcal{O}(\nTx FT_g(\nTx+2))$. For the whole network, the complexity is $\mathcal{O}(CG \nTx FT_g(\nTx+2))=\mathcal{O}(N FT_g(\nTx+2))$, since $CG\nTx=N$.

\section{Performance Evaluation}    \label{sec:PerformanceEvaluation}

\subsection{Scenario and Parameters}  \label{subsec:ScenarioAndParameters}
\begin{table}[tbp]
	\centering
	\caption{System Simulation Parameters}
	\label{table:Simulation_Parameters}
	\renewcommand{\arraystretch}{1.3}
	\begin{tabular}{ll}
		\hline
		Parameter & Value \\
		\hline
		Duplex mode & Half-Duplex\\
		ACIR model & 3GPP mask~\cite{36.942} \\
		$\gammaT$ & $7$ dB \\
		$\Pmax$ & $24$ dBm \\
		$\text{PL}_0$ & $63.3$ dB \\
		$\alpha$ & $1.77$  \\
		$d_0$ & 10 m \\
		$\sigma_1$ & $3.1$ dB \\
		Penetration Loss & $10$ dB per obstructing VUE \\
		$\sigma^2$ & $-95.2$ dBm\\
		$\delta$ & $1/100$ \\
		\davg & 48.6 m \\
		$d_{\text{min}}$ & 10 m \\
		$\beta$ & 0.1 \\
		$\zeta$ & $\gammaT(N\Pmax+\sigma^2) $ \\
		$t^\mathrm{p}$ & 1 \\
		\hline
	\end{tabular}
\end{table}

The simulation parameters are summarized in Table~\ref{table:Simulation_Parameters}. For the ease of reproducibility, we evaluate the proposed algorithms on a fairly simplistic network topology where VUEs are distributed on a convoy. Of course, the proposed algorithms do not assume any particular network topology or simulation parameter values. 

The distance $d$ between any two adjacent VUEs is modeled as a shifted exponential distributed random variable, with minimum distance $d_{\text{min}} = 10\,$m, and average distance $d_{\text{avg}}$ \cite{Koufos1,Dewen1,Cowan1,Luttinen1}. That is, in each trial of the simulation, we drop VUEs in a convoy with random adjacent vehicular distances $d$, whose probability density function is given as,
\begin{equation}
f(d) =
\begin{cases}
(d_{\text{avg}}-d_{\text{min}})^{-1}  \exp({-\frac{d-d_{\text{min}}}{d_{\text{avg}}-d_{\text{min}}}}) , & d \ge d_{\text{min}} \\
0, & \text{otherwise}    
\end{cases}
\end{equation}
where $d_{\text{avg}} = 48.6$\,m, corresponding to 2.5\,seconds for a vehicular speed of $70\,$km/h, as recommended by 3GPP \cite[section A.1.2]{36.885} for freeway scenario. 

We assume that each VUE wants to broadcast its message within $T$ timeslots to the nearest $\nRx$ VUEs, i.e., $\Rs_i$ is the closest $\nRx$ VUEs to VUE $i$. This is in line with \gls{CAMs} scenario proposed by \gls{ETSI}, where the message generation is periodic with periodicity $T$. Furthermore, we set $t^\mathrm{p}=1$, so that the relaying can be done 1 timeslot after the reception. However, note that these $T$ timeslots are allocated to $G$ groups in non-overlapping manner (i.e., each group gets approximately $T/G$ timeslots), and group $g$ gets timeslots $\Ss_g$ as computed in \eqref{compute:Sg}.



\gls{ETSI} defines V2V platooning scenario having message payloads of 300-400 bytes \cite{3gpp22.186}, and the spectrum available for transmission as 5.875--5.905\,GHz \cite[Table 4.2-1]{3gpp303.613}. The physical layer transmission procedures for V2V sidelink is explained in \cite[Section 14]{36.213}. For simulation purpose, we choose a bandwidth of 10\,MHz, which corresponds to 50 RBs in a timeslot. In order to support a message payload of 400 bytes, we set $\gammaT = 7$\,dB, and a VUE is allocated with a contiguous RB-group of 10\,RBs each. Indeed, transmitting in 10\,RBs with SINR 7\,dB achieves sufficiently low error probability for 400 bytes payload. Therefore, in this context, we set the unit of scheduling as RB-Group (consisting of 10 RBs) instead of 1 RB, i.e., we schedule VUEs on each RB-Group instead of RB. This in turn reduces the computational complexity since $F=5$ instead of 50. Indeed, 3GPP support CSI report and scheduling on RB-groups instead of individual RBs to reduce control overhead.

The channel model and parameters are adopted from \cite{Karedal}, which is a model based on V2V link measurements at carrier frequency 5.2 GHz in a highway scenario, and in line with the measurements done in \cite{Abbas1,Lin1,Kunisch1}. The pathloss model is,
\begin{equation}
\text{PL}(d) = \text{PL}_0 + 10 \alpha\hspace*{0.05cm}\text{log}_{10}(d/d_0) + X_{\sigma_1}
\end{equation}
where $d$ is the distance, $\alpha$ is the pathloss exponent, $\text{PL}_0$ is the pathloss at a reference distance $d_0 = 10\,$m, and $X_{\sigma_1}$ is the shadowing effect modeled as a zero-mean Gaussian random variable with standard deviation $\sigma_1$. The penetration loss caused by a blocking vehicle has been widely measured and observed to be 12-13\,dB for a truck \cite{Taimoor}, 15-20\,dB for a bus \cite{He1}, 20\,dB for a van \cite{Meireles1}, and 10\,dB for a car \cite{Abbas2}.  However, there is a lack of enough measurements for the penetration loss caused by multiple obstructing vehicles. Measurements in \cite{Nilsson2} shows that the variance of the shadow fading for two blocking VUEs is greater than for one blocking VUE. For the simulation purpose, we assume penetration loss of $10$\,dB for each obstructing VUE, which might be an over-estimate for the penetration loss. The noise variance is $-95.2\,$dBm and $\Pmax$ is $24\,$dBm as per 3GPP recommendations \cite{36.942}. The $\delta = 0.01$, which implies that the  the worst case inter-cluster interference is limited to 1\% of the noise power. We note that this value of $\delta$ results in $11 \leq d^\mathrm{intr} \leq 13$, consequently, $G=3$ for $\nTx=10$, and $G=2$ for $\nTx \geq 20$.

Finally, the ACIR value $\acir_{r}$ is chosen as the mask specified by 3GPP \cite{36.942}, as follows,
\begin{align}
\acir_{r} &=
\begin{cases}
1, & r=0 \\
10^{-3}, & 1 \leq r \leq 4 \\
10^{-4.5}, & \text{otherwise}          
\end{cases}.
\end{align}

\subsection{Simulation Results}

\newcommand{\folderName}{Plots}

\pgfplotscreateplotcyclelist{colorList1}{
	{black,mark=+},
	{violet,mark=square,mark options={fill=none}},
	{green,mark=triangle},
	{blue,mark=o},
	{red,mark=star},
}

\begingroup
\thickmuskip=0mu		

\begin{figure*}[!ht] 	
	\centering	
	
	\begin{tikzpicture}
	\begin{groupplot}[cycle list name=colorList1, xmajorgrids, ymajorgrids,
	group style={group name=my plots,group size= 3 by 1,vertical sep=2.5cm },
	height=5cm,width=0.3\paperwidth	
	]

	\nextgroupplot[
	xlabel=$\nTx$,
	xtick =  {10, 20, 30, 40, 50} 
	]
	\addplot +[restrict expr to domain={\coordindex}{0:9}]   table [x=xValues, y=sch_Peng1, col sep=comma] {\folderName/nSuccessfullLinks_nTx.csv}; \label{figA:plot1} 
	\addplot +[restrict expr to domain={\coordindex}{0:9}]   table [x=xValues, y=sch_Heuristic1, col sep=comma] {\folderName/nSuccessfullLinks_nTx.csv}; \label{figA:plot2} 
	\addplot +[restrict expr to domain={\coordindex}{0:9}]   table [x=xValues, y=DCS, col sep=comma] {\folderName/nSuccessfullLinks_nTx.csv}; \label{figA:plot3} 
	\addplot +[restrict expr to domain={\coordindex}{0:9}]   table [x=xValues, y=joint_noMultihop, col sep=comma] {\folderName/nSuccessfullLinks_nTx.csv}; \label{figA:plot4} 
	\addplot +[restrict expr to domain={\coordindex}{0:9}]   table [x=xValues, y=joint_basic2, col sep=comma] {\folderName/nSuccessfullLinks_nTx.csv}; \label{figA:plot5} 
	
	\nextgroupplot[
	xlabel=$T$,
	xtick =  {4,8,12,16},
	legend style={at={(0.5,1.08)},anchor=south, legend columns=-1, column sep = 0.1cm}, 
	]
	\addplot  table [x=xValues, y=sch_Peng1, col sep=comma] {\folderName/nSuccessfullLinks_omega.csv}; 
	\addplot  table [x=xValues, y=sch_Heuristic1, col sep=comma] {\folderName/nSuccessfullLinks_omega.csv}; 
	\addplot  table [x=xValues, y=DCS, col sep=comma] {\folderName/nSuccessfullLinks_omega.csv};
	\addplot  table [x=xValues, y=joint_noMultihop, col sep=comma] {\folderName/nSuccessfullLinks_omega.csv};
	\addplot  table [x=xValues, y=joint_basic2, col sep=comma] {\folderName/nSuccessfullLinks_omega.csv};
	\coordinate (top) at (rel axis cs:0,1);
	
	\nextgroupplot[
	xlabel=$\nRx$,
	xtick =  {4,8,12,16,20},
	]
	\addplot  table [x=xValues, y=sch_Peng1, col sep=comma] {\folderName/nSuccessfullLinks_nRx.csv};
	\addplot  table [x=xValues, y=sch_Heuristic1, col sep=comma] {\folderName/nSuccessfullLinks_nRx.csv};
	\addplot  table [x=xValues, y=DCS, col sep=comma] {\folderName/nSuccessfullLinks_nRx.csv};
	\addplot  table [x=xValues, y=joint_noMultihop, col sep=comma] {\folderName/nSuccessfullLinks_nRx.csv};
	\addplot  table [x=xValues, y=joint_basic2, col sep=comma] {\folderName/nSuccessfullLinks_nRx.csv};

	\coordinate[right=-5.7cm] (bot) at (rel axis cs:1,0);
	\end{groupplot}
	\node[below = 1cm of my plots c1r1.south] {(a) ($T=12,\,\nRx=20$)};
	\node[below = 1cm of my plots c2r1.south] {(b) ($\nTx=20,\,\nRx=20$)};
	\node[below = 1cm of my plots c3r1.south] {(c) ($\nTx=20,\,T=12$)};

	\path (top-|current bounding box.west)--
	node[anchor=south,rotate=90] {\parbox{4.5cm}{\centering Average connectivity of a VUE}}
	(bot-|current bounding box.west);
	\path[yshift=2cm] (top|-current bounding box.north)--
	coordinate(legendpos)
	(bot|-current bounding box.north);
	\matrix[
	matrix of nodes,
	anchor=south,
	draw,
	inner sep=0.2em,
	draw
	]at([yshift=1ex]legendpos)
	{
		\ref{figA:plot5} \eqref{formulation:maximizeConnectivity}  ~~
		\ref{figA:plot4} \eqref{formulation:maximizeConnectivity} without multihop  ~~
		\ref{figA:plot3} CDS (Alg. \ref{Alg:CDS}) \\
		\ref{figA:plot2} Heuristic scheduling  \cite{Anver_Access1} ~~~
		\ref{figA:plot1} \cite{Peng1} \\};
	\end{tikzpicture}

	\caption{Average connectivity of a VUE for various algorithms} \label{Fig:nSuccessfullLinks}
\end{figure*}
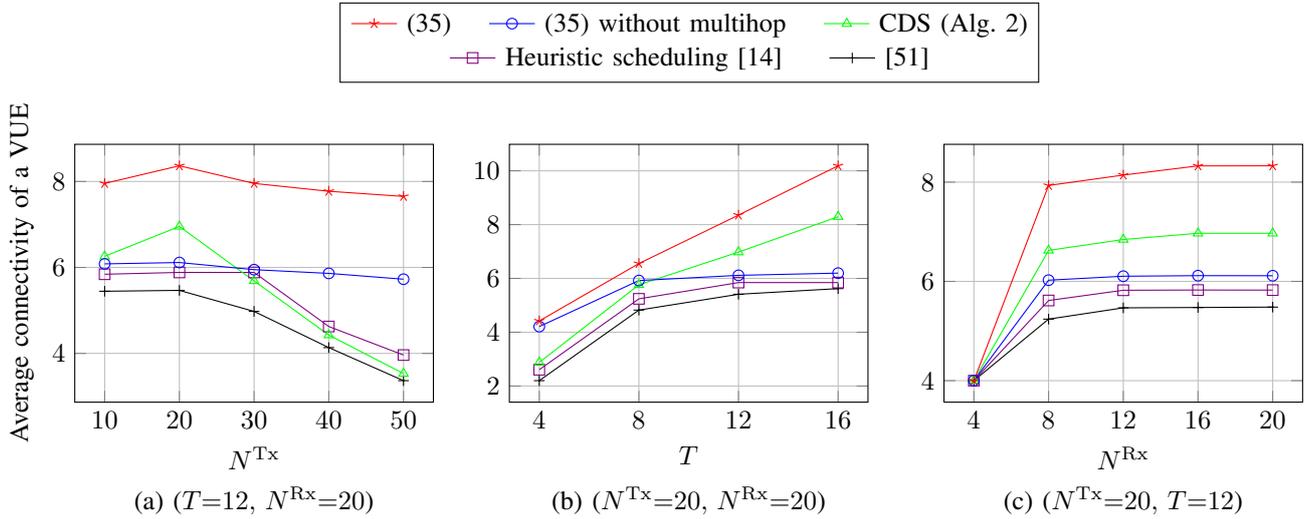

\endgroup




\begingroup
\thickmuskip=0mu		

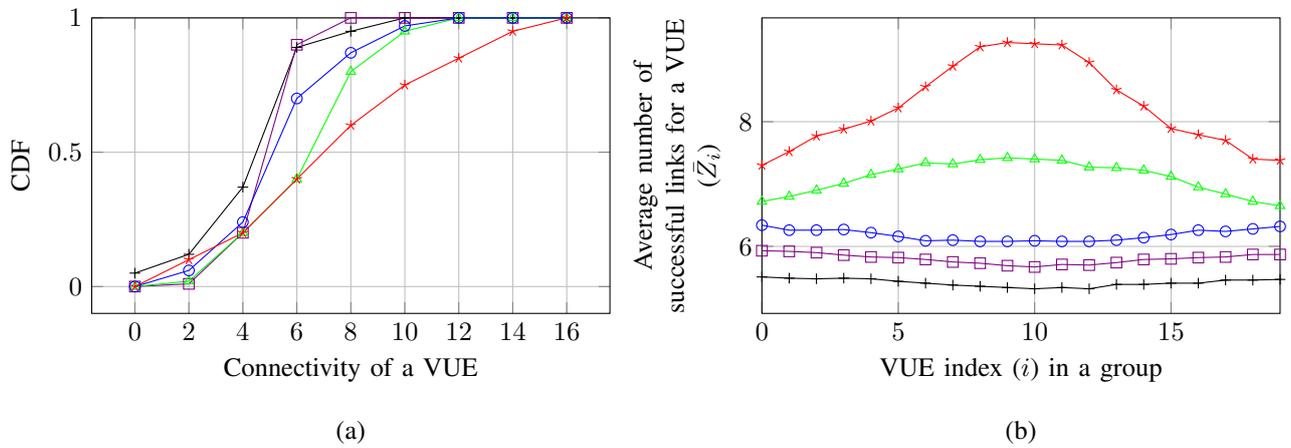
\begin{figure*}[!ht]	
	\centering	
	
	\begin{tikzpicture}
	\begin{groupplot}[cycle list name=colorList1, xmajorgrids, ymajorgrids,
	group style={group name=my plots,group size= 2 by 1,vertical sep=2.5cm ,horizontal sep=2cm },
	height=5.5cm,width=0.4\paperwidth	
	]

	\nextgroupplot[
	xtick =  {0,2,4,6,8,10,12,14,16,18,20},			
	xlabel=\parbox{4cm}{\centering Connectivity of a VUE},
	ylabel=CDF,
	ymax = 1,
	]
	\addplot table [x=xValues, y=sch_Peng1, col sep=comma] {\folderName/CDF_of_nSuccessfullLinks.csv}; \label{figB:plot1} 
	\addplot table [x=xValues, y=sch_Heuristic1, col sep=comma] {\folderName/CDF_of_nSuccessfullLinks.csv}; \label{figB:plot2} 
	\addplot table [x=xValues, y=DCS, col sep=comma] {\folderName/CDF_of_nSuccessfullLinks.csv}; \label{figB:plot3} 
	\addplot table [x=xValues, y=joint_noMultihop, col sep=comma] {\folderName/CDF_of_nSuccessfullLinks.csv}; \label{figB:plot4} 
	\addplot table [x=xValues, y=joint_basic2, col sep=comma] {\folderName/CDF_of_nSuccessfullLinks.csv}; \label{figB:plot5} 
	
	\nextgroupplot[
	xmin = 0, xmax = 19,
	xlabel=VUE index ($i$) in a group,	
	ylabel=\parbox{4cm}{\centering Average number of successful links for a VUE ($\bar{Z}_i$) },
	legend style={at={(-0.15,1.08)},anchor=south, legend columns=-1, column sep = 0.1cm}, 
	]
	\addplot table [x=xValues, y=sch_Peng1, col sep=comma] {\folderName/eachVUE_nSuccessfullLinks.csv};
	\addplot table [x=xValues, y=sch_Heuristic1, col sep=comma] {\folderName/eachVUE_nSuccessfullLinks.csv};
	\addplot table [x=xValues, y=DCS, col sep=comma] {\folderName/eachVUE_nSuccessfullLinks.csv};
	\addplot table [x=xValues, y=joint_noMultihop, col sep=comma] {\folderName/eachVUE_nSuccessfullLinks.csv};
	\addplot table [x=xValues, y=joint_basic2, col sep=comma] {\folderName/eachVUE_nSuccessfullLinks.csv};

	%
	
	\coordinate (bot) at (rel axis cs:1,0);
	\end{groupplot}
	\node[below = 1.3cm of my plots c1r1.south] {(a)};
	\node[below = 1.3cm of my plots c2r1.south] {(b)};

	\end{tikzpicture}

	\caption{Fairness comparison of connectivity of a VUE ($\nTx=20,\,T=12,\nRx=20$)}	 \label{Fig:CDFandEachVUEPerformance}
\end{figure*}

\endgroup

\begin{table*}[t]
	\centering
	\caption{Summary of compared algorithms }
	\label{table:compared_algorithms}
	\renewcommand{\arraystretch}{1.3}
	\begin{tabular}{lllll}
		\hline
		Line Style & Scheduling & Power & Complexity (worst-case) & Algorithm \\
		\hline
		\ref{figA:plot5} & Optimized & Optimized &  $\mathcal{O}(CG\frac{(|\Pb+\Zb|^3 2^{|\Xb|+|\Yb|+|\Wb|})}{\log |\Pb+\Zb|})$ & \eqref{formulation:maximizeConnectivity} \\
		\ref{figA:plot4} & Optimized & Optimized &  $\mathcal{O}(CG\frac{(|\Pb+\Zb|^3 2^{|\Xb|+|\Yb|})}{\log |\Pb+\Zb|})$ & \eqref{formulation:maximizeConnectivity} without multihop \\
		
		\ref{figA:plot3} & Optimized & Equal & $\mathcal{O}(N FT_g(\nTx+2))$ & CDS (Alg.\,\ref{Alg:CDS})\\
		\ref{figA:plot2} & Optimized & Equal & $\mathcal{O}(NFT_g(FT_g+(\nTx)^2))$ & \cite{Anver_Access1}\\
		\ref{figA:plot1} & Optimized & Equal & $\mathcal{O}(N \nTx FT_g)$ & \cite{Peng1}\\
		\hline
	\end{tabular}
\end{table*}

Ideally, we want to consider a vehicular network of a very large size. However, as already mentioned in Section \ref{sec:Clustering}, interference is negligible for clusters beyond \second~neighboring cluster on each side. Therefore, we simulate a network of 5 clusters (i.e., $C=5$), but analyze the performance of VUEs in the the middle cluster (i.e., \third cluster) alone. This is to avoid edge effects, since clusters 1,2,4,5, have less neighbors on one side, hence, unfair to compare.

We set the value of $\nTx$ first, then compute the value of $G$ using~\eqref{compute:G}, and $N = C G \nTx = 5G\nTx$ since the number of clusters is fixed to 5. Recall that the number of groups $G$ depends upon $\nTx$ as per \eqref{compute:G}.

Since the connectivity among VUEs is more important in V2V safety related communication \cite{3gpp22.186}, we show the simulation results for maximizing connectivity by solving \eqref{formulation:maximizeConnectivity}, i.e., maximizing the number of receivers that can successfully receive a message from a VUE. It is worth mentioning that in this scenario, maximizing the connectivity within $T$ timeslots is equivalent to maximizing throughput, since each VUE has a message to multicast in $T$ timeslots. That is, the problem formulations \eqref{formulation:maximizeConnectivity} and \eqref{formulation:maximizeThroughput} are equivalent for our simulation scenario. 

\figref{Fig:nSuccessfullLinks} and \ref{Fig:CDFandEachVUEPerformance} show the simulation results, and the compared algorithms are summarized in Table~\ref{table:compared_algorithms}. As a baseline method, we show the results of the ACI-aware heuristic scheduling algorithm proposed in our previous work \cite{Anver_Access1}, and ACI-unaware multicast scheduling algorithm from \cite{Peng1}. To the best of out knowledge, \cite{Anver_Access1} is the only existing study upon scheduling for maximizing connectivity in V2V multicast communication. The multicast scheduling proposed in \cite{Peng1} is for maximizing the \gls{QoS}, hence, we modify its objective to maximize the connectivity, and present the results. Furthermore, the scheduling algorithms in \cite{Anver_Access1}, \cite{Peng1} are centralized algorithms and require \gls{CSI} between any pair of VUEs in the network, while CDS is a distributed algorithm without the need of \gls{CSI}. The algorithms in \cite{Anver_Access1}, \cite{Peng1} and CDS have polynomial computational complexity, whereas the joint scheduling and power control problem formulations (i.e., \eqref{formulation:maximizeThroughput}--\eqref{formulation:maximizeConnectivity:latency}) have exponential computational complexity. In order to solve all MBLP and BLP problems, we use Gurobi toolbox \cite{Gurobi}.

To quantify the gain due to the multihop, we also show the results for joint scheduling and power control after disabling multihop as blue curves with circles in \figref{Fig:nSuccessfullLinks} and \ref{Fig:CDFandEachVUEPerformance}. The performance gap between \eqref{formulation:maximizeConnectivity} with and without multihop shows the significant improvement due to multihop.

In \figref{Fig:nSuccessfullLinks}\,(a), we plot the average connectivity of a VUE (i.e., $1/|\Ns| \sum_{i \in \Ns} \sum_{j \in \Rs_i}  Z_{i, j}$) for various values of group sizes $\nTx$. The performance improvement for CDS is significant when $\nTx\leq 20$ since the scheduler has more number of RBs to schedule compared to the number of VUEs (i.e., $FT > \nTx$), hence can utilize the extra RBs for multihop to enhance the connectivity. For higher values of $\nTx$, the performance decreases for CDS, \cite{Anver_Access1} and \cite{Peng1} mainly due to their non-overlapping scheduling nature, i.e., an RB cannot be scheduled to more than one VUE.

As we increase the time-horizon for scheduling $T$, the performances of all the algorithms improve as seen from \figref{Fig:nSuccessfullLinks}\,(b). This is not surprising, since more number of timeslots become available for scheduling for each group as we increase $T$. However, for the scheduling algorithms not supporting multihop, the performance do not improve for higher values of $T$, since links beyond \third neighbor on each side of the transmitting VUE tend to be noise limited, due to the high penetration loss of intermediate VUEs \cite{Abbas2}. \figref{Fig:nSuccessfullLinks}\,(c) shows the performance for various number of neighbors to communicate $\nRx$, and one can infer that VUEs are connected more to the close-by neighbors rather than far-away neighbors since the curves flatten out after certain $\nRx$.

To compare the fairness of the schemes, we plot the CDF for the connectivity of a VUE in \figref{Fig:CDFandEachVUEPerformance}\,(a). The high slopes of the CDF show that the simple scheduling algorithms achieve better fairness compared to more advanced \gls{RRM} schemes. Also, note that it is also possible to explicitly enforce fairness, as explained in Section \ref{subsec:ProblemFormulations} variant 3). Similarly, we plot the average connectivity of each VUE in a group of 20 VUEs (i.e., $\nTx=20$) in \figref{Fig:CDFandEachVUEPerformance}\,(b). Note that for algorithms supporting multihop, the connectivity is higher for VUEs in the middle of the group, since they have got more chances for multihoping within the group. We limit multihoping to within a group for the simulation purpose. However, for the algorithms in \cite{Anver_Access1} and \cite{Peng1}, the performance is improved for edge users in the group. This can be due to the fact that edge VUEs can transmit more often to VUEs in the neighboring groups since those VUEs are not transmitting. On the other hand, VUEs within the group are transmitting themselves, hence have less chance for reception due to the half duplex criteria.

It is also worth mentioning that the performance loss due to clustering of the network is subject to $G, \nTx, T$ and multihop nature of the \gls{RRM} schemes. If there are sufficient number of timeslots and multihop is not supported, then a VUE connectivity is saturated to approximately 6 neighbouring VUEs due to the noise limitations. Hence clustering will not affect the performance for no-multihop \gls{RRM} schemes when there are sufficient number of timeslots. However, for multihop \gls{RRM} schemes, the performance improves almost linearly with respect to $T$ (see \figref{Fig:nSuccessfullLinks}\,(b)), however, increasing the size of the network worsen the connectivity marginally only (see \figref{Fig:nSuccessfullLinks}\,(a)). Since clustering effectively reduces the number of timeslots available for a VUE transmission, the performance loss can be significant. Our simulations show that splitting a network having 40 VUEs into two groups with each group having 20 VUEs, reduces the average VUE connectivity from 11.24 to 8.13, when $T=12$. Hence a clustering approach is recommended mainly for the scalability of the network, i.e., to reduce the computational complexity or handle the case when the network controller is absent for the whole network.

\section{Conclusions} \label{sec:Conclusions}

This paper studies the multihop scheduling and power control performance of direct V2V multicast communication in the presence of CCI and ACI. 
From the study and results presented in this paper, we can draw the following conclusions,
\begin{enumerate}
	\item The joint multihop scheduling and power control problem to maximize throughput/connectivity can be formulated as an \gls{MBLP} problem. From this problem formulation, we can derive a scheduling-alone algorithm as a \gls{BLP} problem and a power-control-alone algorithm as an MBLP problem. Similar problem formulation can be done to maximize worst-case throughput/connectivity as well.
	\item To maximize connectivity with a required \gls{AOI}/latency, the joint multihop scheduling and power control can be formulated as an MBLP problem.
	\item The scalability issues of \gls{RRM} schemes can be solved by splitting large networks into smaller clusters, and further splitting each cluster into smaller groups. Inter-group interference within a cluster can be avoided by allocating distinct timeslots to different groups and inter-cluster interference can be made to significantly low by appropriately choosing the cluster size. Each group can schedule and power control independently in its allocated timeslots, thereby, reducing the computational complexity.
	\item In general the algorithms supporting multihop show significant performance improvement in maximizing the connectivity among vehicles.
	\item The proposed CDS algorithm shows improved performance and works in a distributed manner without the need for channel knowledge.
\end{enumerate}

\begin{appendices}

	\section{A Mathematical Background} \label{Appendix:MathematicalBackground}
	As already mentioned, we are trying to formulate all the problems into \gls{MBLP} problems. However, we need to use nonlinear operations, like Boolean \textit{OR}, \textit{AND} and $\min$ operations. Therefore, in this appendix, we explain conversion of \textit{OR}, \textit{AND}, and $\min$ operations into linear constraints, and the whole paper assumes this conversion. 
	
	\subsection{Converting \textit{OR} operation into linear constraints}
	Let $x_1,x_2, \dots x_n$ be Boolean variables. Let $y=x_1 \vee x_2 \vee x_3 \cdots \vee x_n$, be the \textit{OR} value of all $x$ values. In other words,
	\begin{equation}
	y = \bigvee_{i=1}^n  x_i  \label{equation:y1}
	\end{equation}
	We can translate the above nonlinear operation into the following linear constraints,
	\begin{subequations} \label{constr:yALL}
		\begin{align} 
		y &\geq  x_i  \qquad \forall\, i\label{constr:y3}   \\
		y &\leq \sum_{i=1}^{n}  x_i \label{constr:y2}   \\
		y &\in \{0,1\}  \label{constr:y1} 
		\end{align}  
	\end{subequations}
	where \eqref{constr:y1} ensures booleanity of $y$, the constraint \eqref{constr:y2} ensures $y=0$ when all $x$ values are 0, and the constraint \eqref{constr:y3} ensure $y=1$ when any of the $x$ values is 1. Therefore, the $y$ variable satisfying all the constraints in \eqref{constr:yALL} satisfies the equation \eqref{equation:y1}.
	
	\subsection{Converting \textit{AND} operation into linear constraints}
	Similarly \textit{AND} operation (denoted by $\wedge$) can be translated into linear constraints. That is,
	\begin{equation}
	y = \underset{i=1}{\overset{n}{\bigwedge}}  x_i  \label{equation:yB1}
	\end{equation}
	can be converted into the following linear constraints,
	\begin{subequations} \label{constr:yBALL}
		\begin{align} 
		y &\leq x_i  \qquad  \forall\, i \label{constr:yB2} \\
		y &\geq  \sum_{i=1}^{n}  x_i  -(n-1) \label{constr:yB3}    \\
		y &\in \{0,1\}  \label{constr:yB1}
		\end{align}  
	\end{subequations}

	\subsection{Converting $\min$ operation into linear constraints} \label{subsec:minTranslation}
	
	Consider the following problem
	\begin{subequations} \label{formulation:minmin1}
	\begin{align}
		\min & ~ y  \label{minimizey}\\
		\mbox{s.t.} ~~&  \nonumber \\
	    & y = \min_i z_i  \label{constr:yEqualsToMinZ}
	\end{align}
	\end{subequations}
	That is, we want to minimize $y$ but at the same time ensure that 
	$y$ is equal to the minimum of $\{z_1, z_2, \ldots, z_n\}$.
    This problem can be translated into
    \begin{subequations}  \label{formulation:minmin2}
	\begin{align}
		\min & ~ y  \\
		\mbox{s.t.} ~~&  \nonumber \\
	    & \sum_{i=1}^n \mathbbm{1}\{y \geq z_i\} \geq 1  \label{constr:check_y_geq_zi}
	\end{align}
	\end{subequations}	
	which can be further translated into the following \gls{MBLP},
	\begin{subequations}  \label{formulation:minmin3}
		\begin{align}
		\min & ~ y  \\
		\mbox{s.t. } &  \nonumber \\
		&y \geq z_i -\zeta (1-w_i)     \quad \forall i  \label{constr:eta2}\\
		&\sum_{i=1}^{n} w_i \geq 1 \label{constr:eta3}  \\
		&w_i \in \{0,1\}   \quad \forall i  \label{constr:wiBoolean} 
		\end{align}
	\end{subequations}
	That is, we want to minimize $y$ but at the same time ensure that $y$ is greater or equal to at least one of the $z_i$ values.
	The auxiliary Boolean variables $w_i$ indicate if the constraint $y \geq z_i$ is satisfied or not, i.e., $w_i=\mathbbm{1}(y \geq z_i)$. Observe that the constraint \eqref{constr:check_y_geq_zi} is equivalent to \eqref{constr:eta2}--\eqref{constr:wiBoolean}. The parameter $\zeta$ is a sufficiently large number to make constraint \eqref{constr:eta2} hold true when $w_i=0$, for all possible values of $z_i$ and $y$. It is not hard to prove that $\zeta = z^\mathrm{max} - z^\mathrm{min}$ is sufficient when the values of $z$ are limited in an interval, i.e., $z_i \in [z^\mathrm{min},z^\mathrm{max}], ~~ \forall\,i$. 
	
	Note that the minimization in the \gls{AOI} problem formulations discussed in Section \ref{subsec:ProblemFormulations} can be reduced to the above problem formulation \eqref{formulation:minmin1}, where constraint \eqref{constr:AOIijt} can be thought as equivalent to \eqref{constr:yEqualsToMinZ}.

\section{Some Practical Considerations}  \label{Appendix:PracticalConsiderations}

\subsection{Supporting Large Message Payloads} \label{Appendix:SupportingLargePayloads}
If a message payload is too big to fit into an RB, then the message has to be fragmented into smaller packets and each packet has to be transmitted in separate RB. Assume that the message $m$ is fragmented into a set of packets $\Ps_m$, and $X_{i,p,f,t} \in \{0,1\}$ indicate if VUE $i$ transmits the packet $p$ in RB $(f,t)$ or not. Then the constraint \eqref{constr:Wjmt} is modified as follows, \\
$W_{j,m,t}=\bigl( \underset{p \in \Ps_m}{\bigwedge}  \underset{i=1}{\overset{N}{\bigvee}}   \underset{f=1}{\overset{F}{\bigvee}}  (X_{i,p,f,t} \wedge Y_{i,j,f,t})\bigr) \wedge (\bigwedge\limits_{t'=0}^{t-1} \neg W_{j,m,t'})$

\subsection{Supporting Very Low Error Requirements}     \label{Appendix:SupportingLowErrorRate}
The standard approach to achieve low packet error probabilities is to use \gls{HARQ}. However, this requires use of acknowledgements, which is cumbersome in broadcast communications and increases latency. For these reasons, we do not consider retransmission schemes in this paper. To achieve low error probabilities, we can use two other approaches: require higher SINR (which comes at the price of shorter 1-hop transmission range) or multiple repeated transmissions of the same message (which comes at the price of increased radio resource use). In the following, we will discuss both options.

For modern modulation and coding schemes, the packet error probability dependency on SINR can be divided into three SINR regions~\cite{Ryan1,Richardson1}:
\begin{enumerate}
    \item Low SINR region where the error probability close to 1
    \item Medium SINR region where error probability decreases rapidly with SINR (also called the waterfall region)
    \item High SINR region where error probability decreases relative slowly with SINR (also called the error-floor region)
\end{enumerate}
Let $\epsilon(\gamma)$ denote the message error probability over one hop with SINR $\gamma$.
Let us consider an end-to-end connection with $h$ hops that are scheduled to respect the SINR threshold $\gammaT$. That is, the hop SINRs $\gamma_1, \gamma_2, \ldots, \gamma_h$ are all greater or equal to $\gammaT$. Since $\epsilon(\gamma)$ is nonincreasing with $\gamma$, 
$\epsilon(\gamma_\ell)\le\epsilon(\gammaT)$.
Assuming hop errors are independent, the end-to-end error probability is
\begin{align}
    \epsilon^{\text{e2e}}
    (\gamma_1, \gamma_2, \ldots, \gamma_h) 
    &= 1-\prod_{\ell = 1}^h (1-\epsilon(\gamma_\ell))\\
    &\le 1- (1-\epsilon(\gammaT))^{h}\\
    &\le 1- (1-\epsilon(\gammaT))^{\nTx}\\
    &\le \nTx\epsilon(\gammaT),
\end{align}
where the inequalities follow since $\epsilon(\gamma_\ell)\le\epsilon(\gammaT)$, $h\le\nTx$ (where $\nTx$ is the number of transmitters that is controlled by the scheduler), and $1-(1-x)^n\le nx$ for $0\le x\le 1,~ n \geq 1$. 

Hence, for a given requirement $\epsilon^{\text{req}}$ on the end-to-end error probability for an arbitrary scheduled path through the network, we can guarantee that 
\begin{equation}
    \epsilon^{\text{e2e}} \le \epsilon^{\text{req}},    
\end{equation}
 if we select $\gammaT$ such that 
\begin{equation}
    \label{eq:gammaT:from:epsilon}
    \gammaT = \min\{\gamma: \epsilon(\gamma)\le \epsilon^{\text{req}}/\nTx\}.    
\end{equation}
Note that this implies that we are using a higher SINR threshold than required when $h<\nTx$. However, if we operate in the waterfall region, the SINR penalty is small for modest $\nTx$. 

In the case increasing the SINR threshold is not attractive (perhaps because we are operating in the error-floor region), we can resort to using repeated transmissions. Suppose we fix $\gammaT$ such that the 1-hop error probability is upper bounded by $\epsilon(\gammaT)$. The end-to-end error probability for scheduled path with $\nTx$ hops is then $\epsilon^{\text{e2e}}\le\nTx \epsilon(\gammaT)$. If errors occur independently, the error probability after $\rho$ repeated transmissions over the end-to-end connection is $(\epsilon^{\text{e2e}})^\rho$. To achieve the error probability $\epsilon^{\text{req}}$, it is therefore enough to use $\rho = \lceil \log(\epsilon^{\text{req}}) / \log(\nTx\epsilon(\gammaT)) \rceil$ repeated transmissions.

To support repeated transmissions, \eqref{constr:Wjmt} has to be replaced by the following set of constraints,
\begin{subequations} \label{constr:retr}
\begin{align}
    \tilde{W}_{j,m,t} &= \bigvee_{i=1}^{N}   \bigvee_{f=1}^{F}  X_{i,m,f,t} \wedge Y_{i,j,f,t}  \label{constr:retr1}\\
    W_{j,m,t} &\leq \rho + 1 - \sum_{t'=0}^{t} \tilde{W}_{j,m,t} + \zeta'(1-W_{j,m,t})  \label{constr:retr2}\\
    W_{j,m,t} &\geq \rho + 1 - \sum_{t'=0}^{t} \tilde{W}_{j,m,t} - \zeta'(1-W_{j,m,t})  \label{constr:retr3}\\
    W_{j,m,t} &\leq \bigwedge_{t'=0}^{t-1} \neg W_{j,m,t'}  \label{constr:retr4}\\
    W_{j,m,t} &\in \{0,1\}
\end{align}
\end{subequations}
where $\tilde{W}_{j,m,t}$ indicate if message $m$ is received by VUE $j$ during timeslot $t$ with 1-hop error probability $\epsilon$. The constraints \eqref{constr:retr2} and \eqref{constr:retr3} are to ensure that  $W_{j,m,t} = 0$, when $\sum_{t'=0}^t \tilde{W}_{j,m,t'} \neq \rho$. The parameter $\zeta'$ is a large number to make constraints hold when $W_{j,m,t}=0$. It is not hard to prove that $\zeta'=T$ is sufficient. The constraint \eqref{constr:retr4} is to ensure that $W_{j,m,t}=1$ only when the message is received for the first time with error probability less than or equals to $\epsilon^\mathrm{req}$. The main drawback with this scheme is that $\rho$ repeated transmissions is used also when $h < \nTx$. This is wasteful, especially for 1-hop ($h=1$) communication.

\section{Age of Information Requirements}
\label{appendix:AoI}

We recall from~\eqref{constr:AOIijt} that $A_{i,j,t}$ can be computed for $t\in\Ss$ as
\begin{equation}
\label{eq:Aijt:app}
A_{i,j,t} = \min_{m \in \Ms_{i}} (t + A^\mathrm{init}_{i,j} + 1-  (\arrivalTime+A^\mathrm{init}_{i,j} + 1) \sum_{t'=0}^{t} W_{j,m,t'}),
\end{equation}
We see that $A_{i,j,t}$ is a deterministic function of $t$ that depends on the scheduling and power allocation through $W_{j,m,t}$. Indeed, where we recall from~\eqref{def:Wjmt} that
\begin{equation}
    W_{j,m,t} =
    \begin{cases}
        1, &
        \parbox[t][][t]{5.5cm}{if message $m$ is \textit{high-SINR scheduled} to VUE $j$ for first time in timeslot $t$}\\
        0, & \text{otherwise}
    \end{cases}
\end{equation}
That is, if $W_{j,m,t'}=1$, then message $m$ is scheduled to be transmitted by some VUE~$i'$ in an RB $(f,t')$ where the received SINR at VUE~$j$ is high: $\gamma_{i', j, f, t'} \ge \gammaT$. Moreover, for all previous transmissions of message $m$, the received SINR at VUE $j$ is less than $\gammaT$.  

However, the true AoI is a random process that depends on which messages that have been delivered error-free. We can find the true AoI, $A_{i,j,t}^\mathrm{E}$, by replacing $W_{j, m, t}$ in \eqref{eq:Aijt:app} with
\begin{equation}
    W_{j,m,t}^\mathrm{E} =
    \begin{cases}
        1, &
        \parbox[t][][t]{5.5cm}{message $m$ is \textit{delivered error-free} to VUE $j$ for first time in timeslot $t$}\\
        0, & \text{otherwise}
    \end{cases}
\end{equation}
The superscript $\mathrm{E}$ is to indicate that the $W_{j,m,t}^\mathrm{E}$ and $A_{i,j,t}^\mathrm{E}$ are random due to transmission errors (which are random). 

In general, $A_{i,j,t}$ is neither an upper nor a lower bound\footnote{To see this, suppose $W_{j, m, t'}=1$. It is possible that the scheduled transmission at $t=t'$ suffers a transmission error, and message $m$ is not delivered error-free at time slot $t'$. Moreover, it is also possible that 
the message $m$ is delivered error-free in timeslot $t=t''<t'$, although the SINR at $t''$ is less than $\gammaT$.}  on  $A_{i,j,t}^\mathrm{E}$. 
Nevertheless, we will show that $A_{i,j,t}$ can be used to design a schedule and power allocation such 
that a probabilistic performance metric on $A_{i,j,t}^\mathrm{E}$ satisfies a predetermined requirement.

As mentioned in Section~\ref{sec:AoI:computation}, 
we will consider probabilistic AoI requirements of the form
\begin{equation}
    \label{eq:AoI:prob:req:repeated}
    \Pr\{\mu(A_{i,j,t}^\mathrm{E}) \le \mu^\mathrm{T}\} \ge P_{A}^{\text{req}}
\end{equation}
where the metric $\mu$ is a mapping from $(A_{i,j,t}^\mathrm{E}: t\in\Ss)$ to $\mathbb{R}$, $\mu^\mathrm{T}$ is the metric threshold, and $P_{A}^{\text{req}}$ is the required probability. The metric $\mu$ is such that if $A_{i,j,t}' \le A_{i,j,t}$ for $t\in\Ss$, then $\mu(A_{i,j,t}') \le \mu(A_{i,j,t})$. 

Now suppose the schedule and power allocation is such that $\mu(A_{i,j,t}) \le \mu^\mathrm{T}$. We will now show that this implies that $\Pr\{\mu(A_{i,j,t}^\mathrm{E}\} \le \mu^\mathrm{T}\}$ is greater than a probability that can be controlled by the SINR threshold. 

From \eqref{eq:Aijt:app}, we see that $A_{i,j,t}$ is determined by $M_{i,j}$ scheduled transmissions, where
\begin{equation}
    M_{i,j} = \sum_{m\in\Ms_i} \sum_{t=0}^{T-1} W_{j,m,t}.
\end{equation}
That is, $M_{i,j}$ is the number of messages that are generated by VUE~$i$ and \textit{high-SINR scheduled} to transmit to VUE~$j$. We note that $M_{i,j}\le |\Ms_i|$ since, for a fixed $m$ and $j$, $\sum_{t=0}^{T-1} W_{j,m,t}\le 1$.
Let $G$ denote the event that all of these \textit{high-SINR scheduled} messages are delivered error-free. Assuming independent end-to-end message errors, we can write 
\begin{equation}
    \Pr\{G\} \ge (1-\epsilon^{\text{req}})^{M_{i,j}}
    \ge (1-\epsilon^{\text{req}})^{|\Ms_i|} 
    \label{eq:PE:bound}
\end{equation}
where the first inequality holds since $\gammaT$ is set sufficiently large to ensure that the end-to-end error probability $\epsilon^{\text{e2e}}\le \epsilon^{\text{req}}$ (see Appendix~\ref{Appendix:PracticalConsiderations}) and the second inequality holds since $M_{i,j}\le |\Ms_i|$.

The crucial observation is that, conditioned on the event $G$, if $W_{j, m, t'} = 1$ for some message $m\in\Ms_i$, then message $m$ is delivered error-free at timeslot $t'$. Hence, $W_{j,m,t''}=1$ for some $t''\le t'$. The inequality is due to the facts that (a) conditioned on $G$, message $m$ is delivered error-free at timeslot $t'$ and (b) it is possible that the message $m$ is transmitted in an RB $(f'', t'')$ and delivered error-free, even though the SINR for this transmission is less than $\gammaT$. (Fact (b) holds regardless if we condition on $G$ or not).    
Now, additional received copies of message $m$ cannot increase the AoI and it follows that, conditioned on $G$, $A_{i,j,t}^\mathrm{E} \le A_{i,j,t}$ and
\begin{equation}
    \mu(A_{i,j,t}^\mathrm{E}) \le \mu(A_{i,j,t}) \le \mu^\mathrm{T}.
    \label{eq:Aijte:bound}
\end{equation}
Hence, if we by $G^\mathrm{c}$ denote the complement of the event $G$, we have that
\begin{align}
    \Pr\{\mu(A_{i,j,t}^\mathrm{E}) \le \mu^\mathrm{T}\} &=
    \Pr\{\mu(A_{i,j,t}^\mathrm{E}) \le \mu^\mathrm{T}\mid G\}\Pr\{G\} \nonumber\\
    &\quad +\Pr\{\mu(A_{i,j,t}^\mathrm{E}) \le \mu^\mathrm{T}\mid G^\mathrm{c}\}\Pr\{G^\mathrm{c}\}\nonumber\\
    &\ge \Pr\{\mu(A_{i,j,t}^\mathrm{E}) \le \mu^\mathrm{T}\mid G\}\Pr\{G\}\label{eq:PA:bound:1}\\
    &= \Pr\{G\}\label{eq:PA:bound:2}\\
    &\ge (1-\epsilon^{\text{req}})^{|\Ms_i|} \label{eq:PA:bound:3}.
\end{align}
where \eqref{eq:PA:bound:1} follows since probabilities are nonnegative, \eqref{eq:PA:bound:2} since $\Pr\{\mu(A_{i,j,t}^\mathrm{E}) \le \mu^\mathrm{T}\mid G\}=1$ due to~\eqref{eq:Aijte:bound},
and \eqref{eq:PA:bound:3} follows from~\eqref{eq:PE:bound}.

We can therefore conclude that the probabilistic requirement~\eqref{eq:AoI:prob:req:repeated} is satisfied if $\mu(A_{i,j,t}) \le\mu^{\mathrm{T}}$ and $(1-\epsilon^{\text{req}})^{|\Ms_i|} \ge P_A^{\text{req}}$.

\end{appendices}

\bibliographystyle{IEEEtran}
\bibliography{MultihopCommunication}	

\end{document}